# Superconducting phase transition reveals an electromagnetic coupling to a scalar field potential that generates mechanical work

Mark Gibbons CEng MEI

Target Carbon Limited, 271 Coppice Road, Poynton, Stockport, Cheshire, SK12 1SP, United Kingdom

Email: markgibbons@targetcarbon.co.uk

Abstract

Pressure-induced, spontaneous diamagnetism associated with critical behaviour is determined experimentally in a polar dielectric fluid containing nanoscale, clathrate hydrate cage structures. As with Type II superconductivity, Abrikosov vortices come to penetrate the external diamagnetic field such that it reduces to zero for particular values of the magnet flux. The external magnetic field is thus revealed to be the order parameter that signifies a phase transition between Type II superconducting behaviour and a dual of Type I superconducting behaviour. This phase transition is described by a distinctive universality class of critical exponents. The Abrikosov vortices are interpreted as effective magnetic monopole defects associated with the non-equilibrium, geometrically frustrated system. The magnitude of the spontaneous Type I response is consistent with exponential coupling of the spontaneous magnetism with an external scalar field potential made accessible through inertia and hyperbolic geometry. Under this interpretation, magnetic monopole defects act as inhomogeneous nucleation sites able to expand or contract the volume of the system in an analogue of cosmological inflation. The quantum vacuum origin of the scalar field is held responsible for the resulting mechanical work, so representing a potentially unlimited source of zero-emissions energy.

Introduction and background

The original objective for this research sought to reduce compression-stage losses associated with thermodynamic cycles. Isochoric thermo-compression of polar dielectric fluids at temperatures below 373K was selected as a suitable area of investigation to reduce the mechanical compression penalties associated with irreversible thermodynamic cycles.

The manipulation of dissipative, reorganizing, clathrate structures [1,2,3,4] within a polar dielectric inhibitor solvent [5] is reported. The resulting crystal-fluid is subjected to tension, or negative pressure, which establishes metastability [6,7,8] and long-range interaction [9]. The inclusion compounds and dielectric solvent together produce a low-energy system capable of forming extensive hydrogen bonding pathways for magnetic exchange interactions [10,11]. The tetrahedral molecular structure of the crystal-fluid represents a geometrically frustrated system from which magnetic effects can arise [12,13,14].

The experimental investigations documented here describe the emergence of spontaneous diamagnetic and paramagnetic behaviours as critical phenomena [15] that are responsible for work output in a kinetic system, ie. one performing variable-volume, mechanical work. In implementing this power cycle, the crystal-fluid is described by a constant Hamiltonian function, ie. it behaves as a linear oscillator [16]. The experimental findings are presented in a series of tables and charts derived from recorded and calculated data. In positing reasonable theories to elucidate the experimental results, it is necessary to move beyond classical thermodynamics to uncover potential solutions in condensed-matter physics, statistical mechanics, topology and quantum mechanics.

The existence of a metastable, liquid–liquid critical point in tetrahedrally coordinated systems, such as water, silica, silicon, and carbon has been postulated by Smallenburg *et al* [17]. According to this hypothesis, two liquid phases differing in density appear at low temperatures or low energies. The existence of such liquid–liquid transitions ending in a critical point has been identified as responsible for many thermodynamic anomalies observed in these systems, including local extrema in density, compressibility, and heat capacity [17,18].

In the vicinity of the spinodal, as identified in **Fig. 1**, the response functions of metastable systems all increase, ie. the behaviour of thermodynamic values is the same as for critical phenomena [19]. Such a system is adequately described by pseudo-critical indices even though the system itself may be far from a critical point. The susceptibility of an unbounded system at spinodal points tends to infinity and, correspondingly, the volumetric integral of the pair correlation function of the density diverges. This means that fluctuations become long-range, and the system effectively approaches pseudo-critical states.

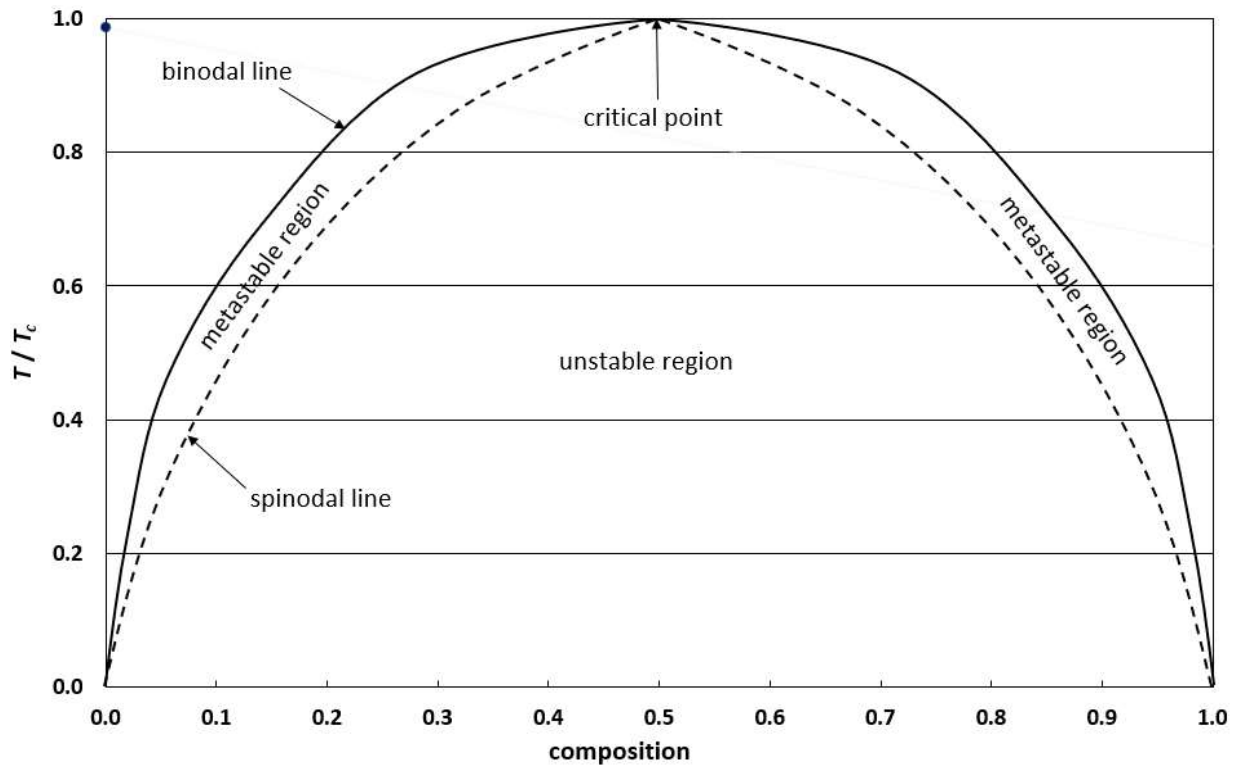

Fig. 1: Liquid-vapour phase separation and spinodal line for a binary system

The presence of long-range correlations within the crystal-fluid is investigated. Statistical field theory provides for a macroscopic, qualitative description of phase transitions and critical phenomena [20,21,22]. For example, the divergence of correlation length in the vicinity of a second-order phase transition suggests that properties near the critical point can be accurately described within an effective theory involving only long-range fluctuations of the system, ie. qualitative properties such as scaling behaviour are completely defined [20].

The modern approach to classifying phase transitions is established within two broad categories. Discontinuous (first-order) transitions incorporate isobaric, latent heat transfers. Continuous (second-order) transitions do not involve latent heat behaviour and are characterized by a steady change in specific volume and entropy [23]. Continuous phase transitions can be associated with an emergence of magnetism, superconductivity, superfluidity and orientational order. Critical properties in the vicinity of continuous phase transitions fall into a limited number of universality classes defined by the fundamental symmetries of the system [15,24]. When a symmetry is broken, a corresponding order parameter that diminishes to zero can often be identified. For critical properties, the scaling of singularities in the order parameter can be determined through the application of critical exponents [25]. The exponents describe the non-analyticity of various thermodynamic functions and are applicable to different categories of phase transitions such that they describe a universality class [15,24].

In crystalline, clathrate hydrate compounds, water molecules link together via hydrogen bonds to form host hydrate lattices, whilst guest hydrocarbon molecules can be retained in the cavities of the framework due to vdW interactions [26]. Through this non-covalent mechanism, water and various gas molecules can form inter-linked cages in which the inner cavities are either entirely or partially filled, or even devoid of guest molecules. A water clathrate without guest molecules constitutes the lowest density ice phase of water. Low-density phases are thermodynamically favoured by low or negative pressures, such that empty clathrates may be a stable crystal of water under tension [27,28].

When behaving as soft supramolecular materials, clathrate hydrates are able respond to changing external conditions through dramatic structural reorganization [16,29,30,31]. The complex reorganization of such dissipative structures can be sustained indefinitely in open systems by a flow of matter and energy. For closed systems, only the exchange of energy is required. Although such organized states are associated with local entropy reduction, entropy is increased globally in accordance with the second law of thermodynamics [32].

For a complex, dissipative system that is far from equilibrium, scale-free fluctuations and correlations drive spontaneous, self-organizing behaviour able to counterbalance a critical response in the absence of a global extremum principle [3]. Under these conditions, the thermodynamic potentials of Gibbs energy, Helmholtz energy, and enthalpy all relate to metastable states that produce only local solutions.

These experimental findings provide qualitative and quantitative evidence for pressure-induced, spontaneous magnetism arising from the magnetic exchange interactions of dissipative, nanoscale, clathrate hydrate cage structures [33,34,35]. Critical phase-change phenomena give rise to transient Type II and Type I superconductor-like behaviours that interact with a magnetocaloric cycle [36,37] to generate mechanical work output. Meeting the requirement for energy conservation when proceeding through a complete work cycle leads to some remarkable results.

Methods and materials

Previous results [9] indicate the presence of an unshielded, long-range vdW interaction responsible for non-equilibrium, phase-change work in a quasi-thermodynamic cycle. The latest experimental investigations were originally designed to extend this vdW interaction potential to generate mechanical work, ie. one producing a variable-volume, kinetic output. Whist the experiment was successful in creating a mechanical work output, this cannot be attributed to the long-range vdW interaction and alternative description has become necessary.

The original apparatus is modified such that the 5 litre pressure vessel is here replaced with a 0.5 litre piston accumulator/ expander, as described in **Fig. 2** below:

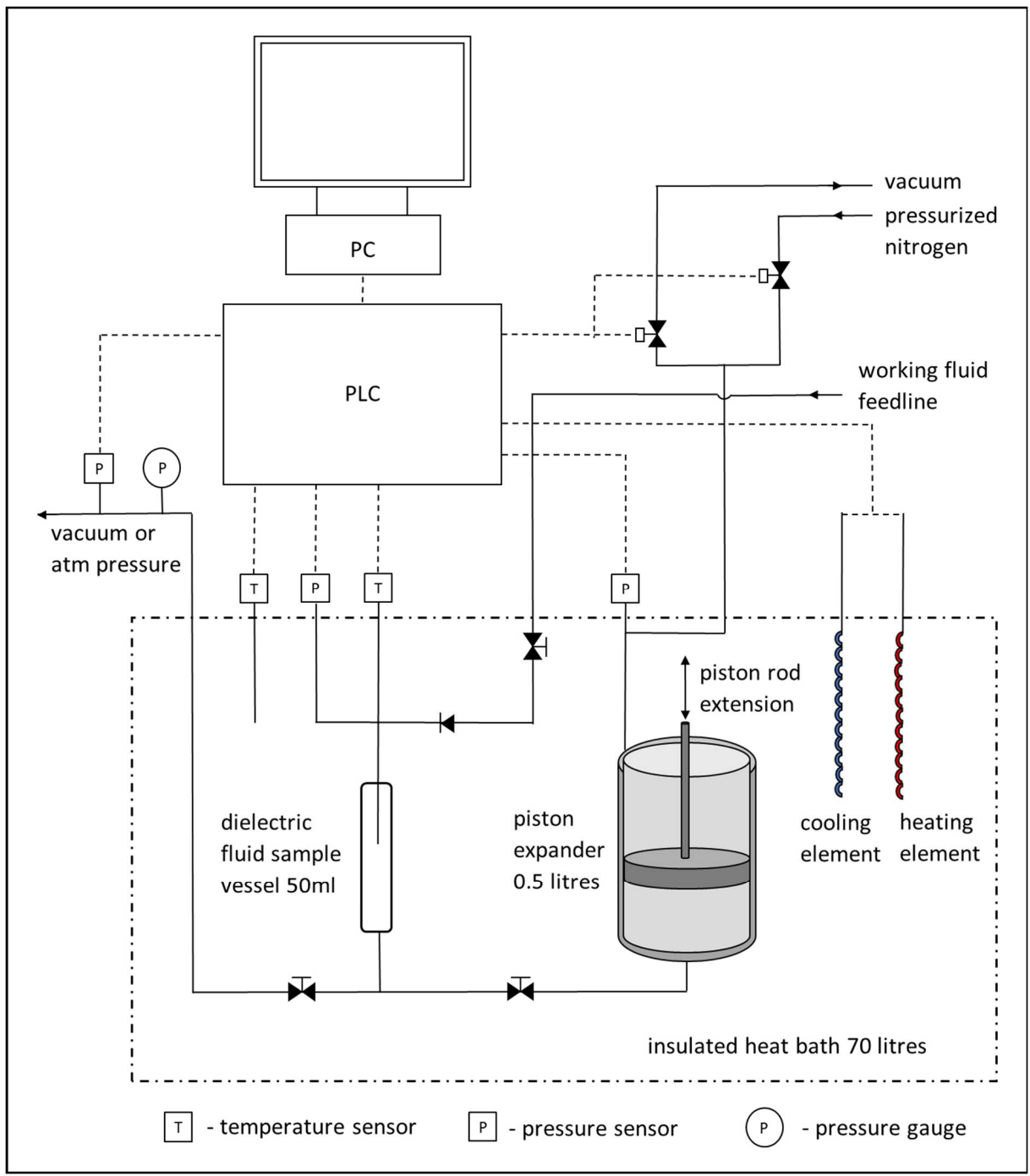

Fig. 2: Schematic arrangement of the experimental apparatus

The current investigations also centre on a polar dielectric working fluid of multi-component formulation. This consists of water and methane within an inhibitor solvent engineered such that the equilibrium between attractive and repulsive molecular forces is readily destabilized when the fluid is subjected to variable pressure. The nature and purpose of the clathrate inhibitor is described in detail by Perrin *et al* [38]. Our latest inhibitor formulation is designed to maximize the mechanical work of a quasi-thermodynamic cycle at relatively low temperatures (ie. below 293K). Changes in the crystal-fluid properties are controlled with a view to establishing a power cycle through non-equilibrium, irreversible volume displacements.

A negative-pressure fluid is once again established by means broadly similar to the Berthelot-method [39]. The fluid, as specially formulated for this application, fills a previously evacuated stainless-steel sample vessel (50ml), as **Fig. 2** above. A low-energy, negative-pressure regime results in the formation of clathrate hydrates hosting methane molecules. The sample vessel is completely immersed in a relatively large heat bath (70 litres) where the temperature of the bath is controlled with an electric element and a refrigeration dip cooler.

Once the desired temperature is obtained, the sample is released into the fluid-side of the 0.5 litre retracted piston expander, also completely immersed in the heat bath, which displaces the piston vertically upwards to the fully-extended position. The gas-side of the piston is open to atmospheric pressure during this extension. This action reduces the energy of the system further and is intended to transfer the guest methane molecules from the host clathrate cages to similar structures within the inhibitor solvent (ie. to maintain the fluid in a liquid phase and avoid the methane outgassing reported in earlier investigations [9]).

The gas-side outlet is then isolated from atmospheric pressure. Through the application of pressurized nitrogen gas, the piston is returned to its fully-retracted position and the valve between the sample vessel and the piston expander closed such that the effective sample is now only 5ml, approx. A salient feature noted here is that the piston displacement action appears unaffected by the status of this isolating valve such that the same results are obtained regardless of the amount of crystal-fluid in direct contact with the piston expander. Additionally, the pressure-temperature instrumentation installed on the side of the sample vessel records values of the same magnitude irrespective of the isolating valve status; an apparent anomaly that will be addressed later in the discussion.

At this point the pressurized nitrogen is released to atmosphere resulting in positive displacement of the piston. Negative and positive piston displacements are induced to produce negative and positive work outputs with an effective compression ratio of 100:1, approx.

The extent of each piston movement is recorded to determine the pressure-volume *PV* work input and output resulting. This cyclic process can be repeated many times without any apparent degradation in *PV* work or the speed of operation. The temperature and pressure of the fluid are measured at five-second intervals with sensors that have direct contact with the fluid (which can be either mechanically connected or isolated from the piston expander) and recorded by a PLC/ PC monitoring system.

Results

The variable-volume results from the piston-expander apparatus are presented below. All values for energy and thermodynamic potentials are derived from the pressure and temperature measurements by the REFPROP program/database [40]. The calculations are in accordance with GERG-2008 modified by the Kunz and Wagner Model 0 (KW0) [41]. Temperature and pressure measurements for a manually-controlled sequence of 10-cycles establish the profiles in **Fig. 3**. The heat bath temperature on this occasion is 270.5K, approx. However, power cycles can also be created with heat bath temperatures up to 288K.

It is apparent that the externally-applied nitrogen pressure does not fully account for the work input or output of the system. Where the fluid is considered to be an ideal gas, then compression action would multiply the pressure at Point 1 by a factor of 100 (ie. from 0.29 MPa to 29 MPa). Correspondingly, expansion action would reduce the pressure at Point 3 by a factor of 100 (ie. from 0.61 MPa to 0.006 MPa). An incompressible liquid would produce much large pressure variations and insignificant volume changes.

Stage 1-2 can therefore be interpreted as negative displacement of the piston (rather than compression of the working fluid) from a positive pressure perturbation. Stage 3-4 is taken to be positive displacement of the piston (rather than expansion of the working fluid) from a negative pressure perturbation. The piston achieves the maximum stroke length for each external perturbation:

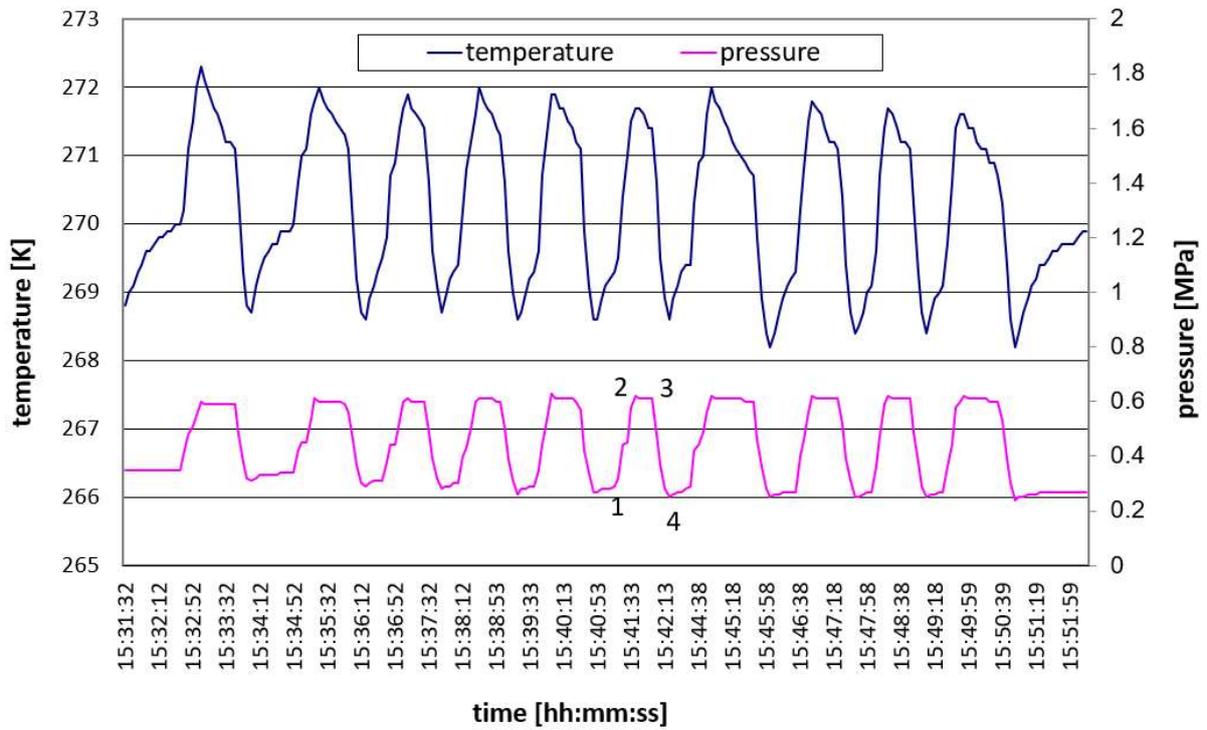

Fig. 3: Temperature and pressure profiles for a sequence of 10-cycles

Calculated thermodynamic properties for Points 1-4 are included in **Appendix A**, **Table A1**.

The heat bath remains at almost constant temperature with no significant heating or cooling during the measurement period. To compare this to a thermodynamic system based upon separate hot and cold reservoirs, the minimum fluid temperature is taken as $T_{min}$ and the maximum fluid temperature is taken as $T_{max}$. At just over 1%, it is clear that the ideal Carnot efficiency ($1 - T_{min}/T_{max}$) cannot adequately describe the non-equilibrium system under investigation.

As with the isochoric results [9], the variable-volume experiments produce a non-concave entropy function of internal energy, indicative of local stability conditions, as shown in **Fig. 4**:

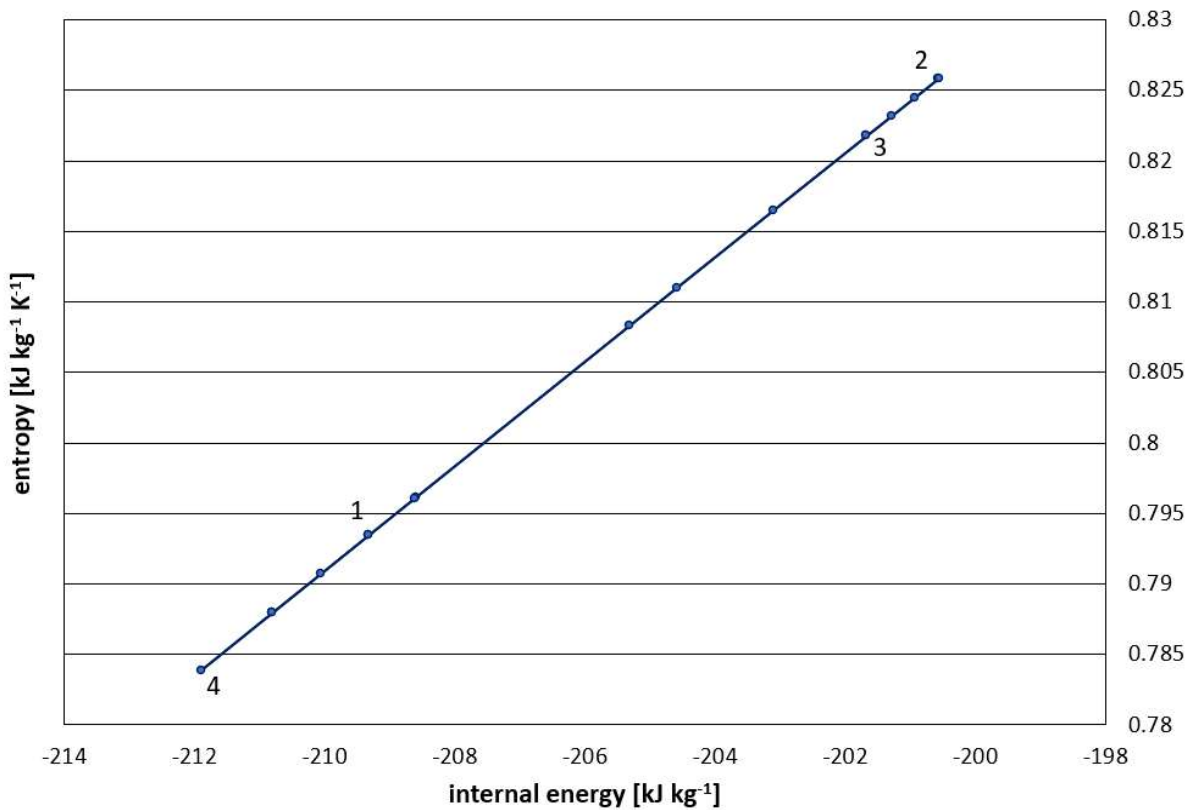

Fig. 4: Affine (non-concave) entropy revealing system inhomogeneities

Relative, negative values of internal energy indicate a low-energy system resulting from tension, or negative pressure. Since enthalpy of the system is dominated by internal energy, a linear enthalpy-entropy compensation effect is also demonstrated [31].

A further correspondence with the original isochoric results [9] comes from the excess negative internal energy arising from a disparity between the thermodynamic potentials and the fundamental thermodynamic relation for the non-equilibrium system, as **Table A1** and plotted in **Fig. 5**. Increasing internal energy is mirrored by an increasing negative excess internal energy potential:

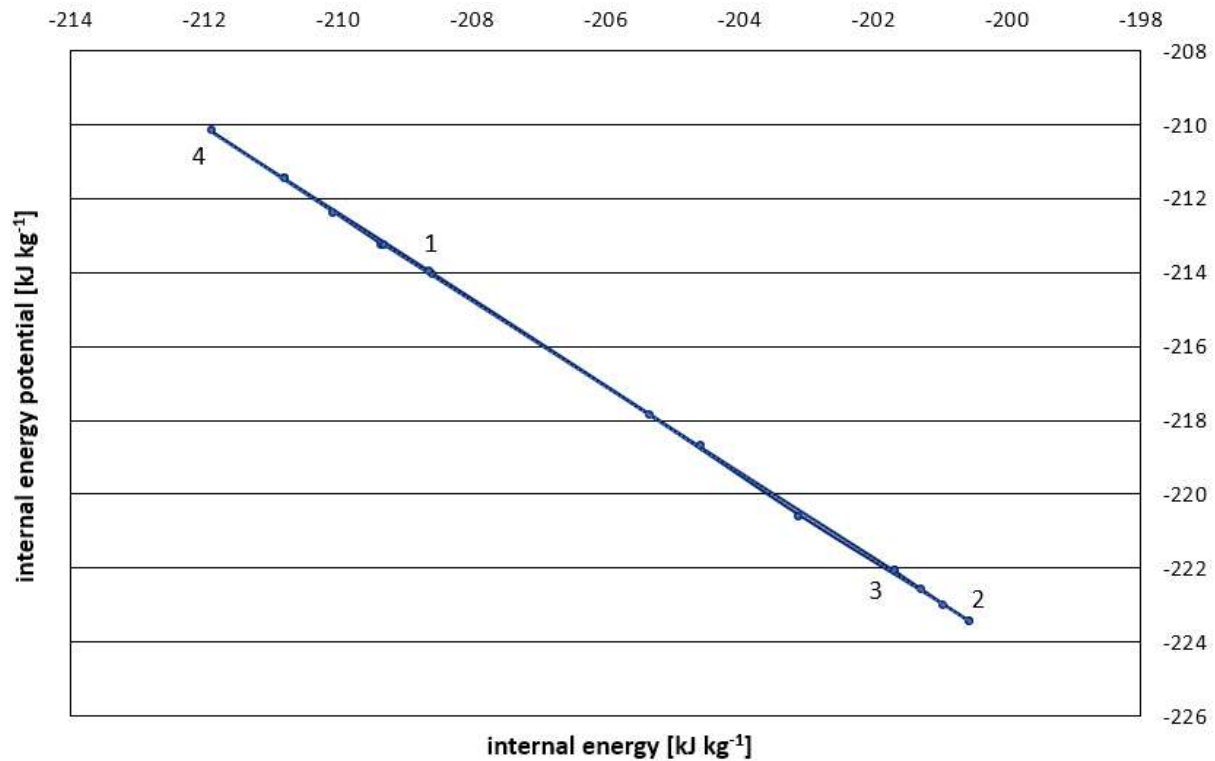

Fig. 5: Excess internal energy potential resulting from excess thermodynamic potentials

Changes in negative potential energy and internal energy vary in a 1:1 relationship after discounting the *Pv* term associated with the walls of the sample vessel. Thus, a linear oscillator or constant Hamiltonian function is described.

When proceeding through a cycle, the calculated density remains almost constant such that it is not possible to generate work output from a first-order phase transition, ie. *Pv* ≠ *PV* where ***V*** is the specific *swept* volume. Additionally, **Table A2** and **Fig. 6** below reveal similar calculated values for work input (Stage 1-2) and work output (Stage 3-4) for the single cycle:

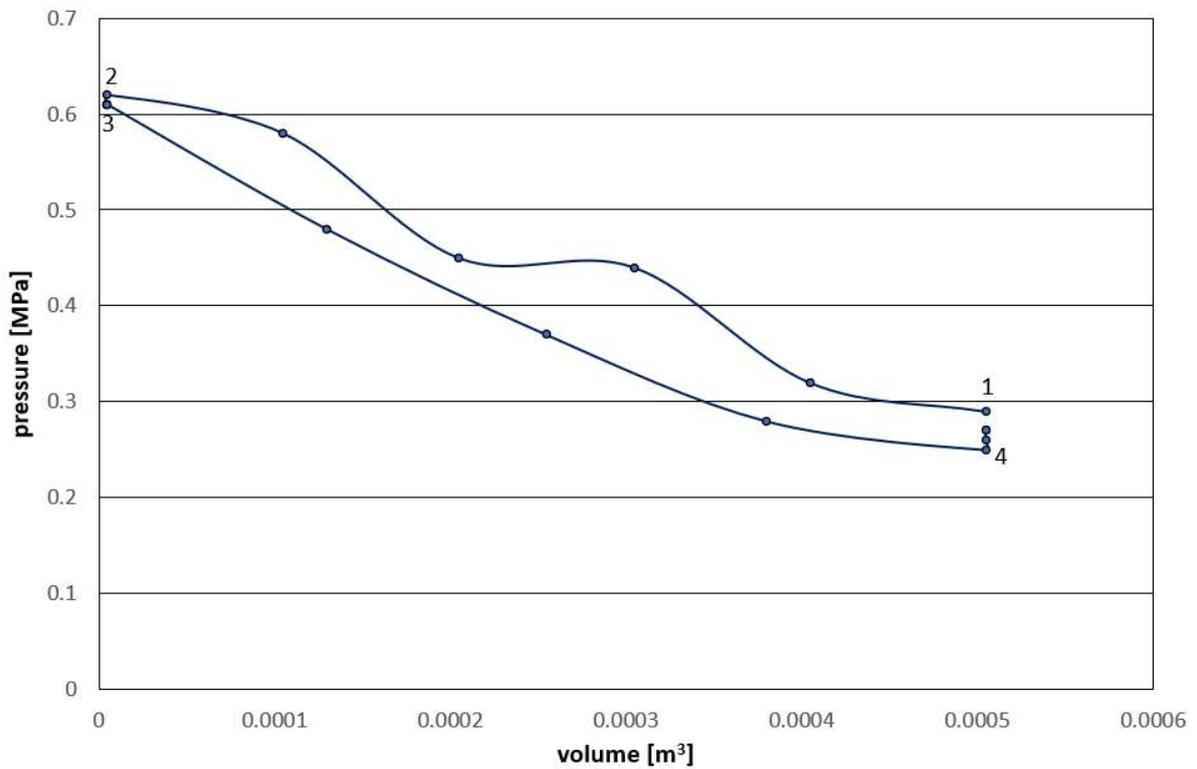

Fig. 6: *PV* work input and *PV* work output can be determined

The disparity between *Pv* and *PV* can be considered in terms of hyperbolic fluid curvature. Relating the potential energy component of the Hamiltonian (ie. the pressure-density term) to system inertia ($I = \tfrac{1}{2}mr^2$) produces a range of values for the Gaussian radius of curvature $R_g$ which can be compared to system pressure *P*, as described in **Fig. 7** and **Appendix B**:

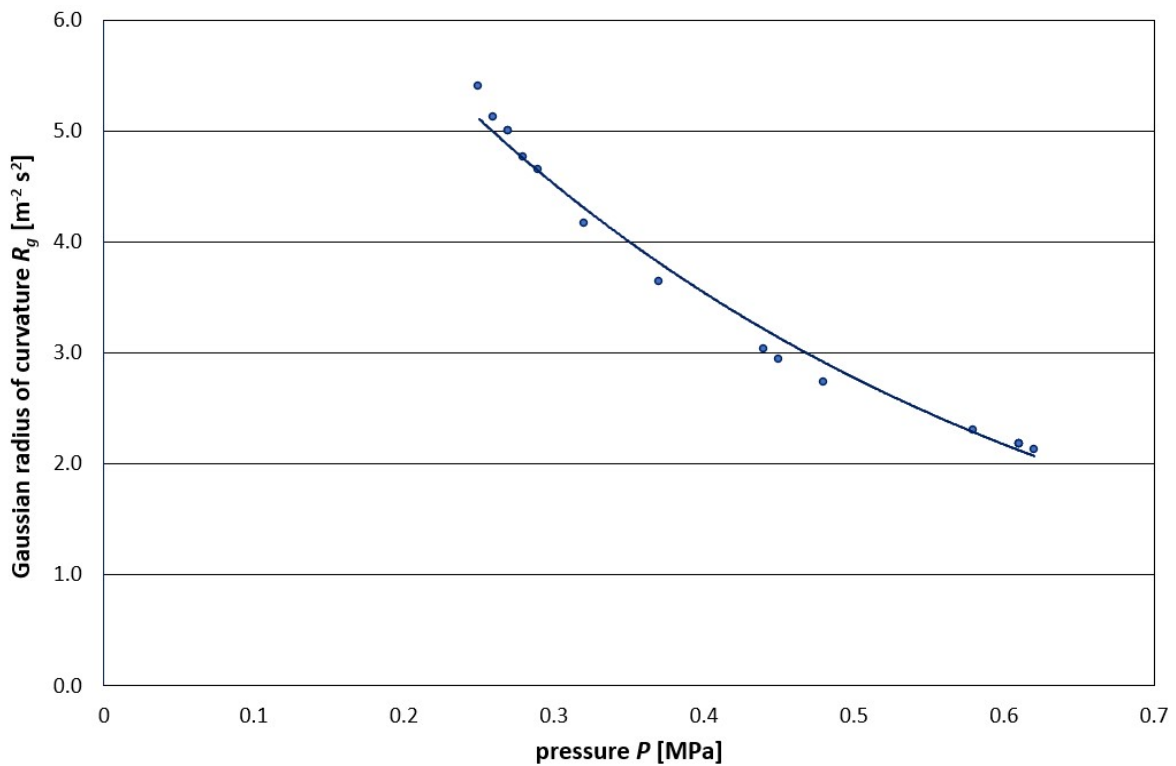

Fig. 7: Non-extensivity of specific volume expressed as curvature

At Points 1-4 it is possible to map a gradient energy term to hyperbolic surface area to reveal an approximate negative-inverse relationship in **Fig. 8**. Values for gradient energy are derived where *PV* work expresses the least action principle of the Langrangian function (**Appendix B**). The hyperbolic surface area of the fluid can be determined from the effective radius through the relation *A* =

$4\pi\sinh^2(R/2)$, where $A$ is the hyperbolic surface area of a hollow pseudo-sphere and $R$ is the effective radius ($r_1 - r_2$) for the hollow pseudo-sphere ($r_1$ is the outer radius and $r_2$ is the inner radius) under the geometrical constraint $0 < r_2/r_1 < 1$ (**Table A3** and **Appendix B**).

Gradient energy and curvature can be directly compared through dimensional analysis. However, during external pressure perturbations (1→2 negative displacement and 3→4 positive displacement) the geometrical constraint produces a disparity in the mapping that corresponds to phase transitions:

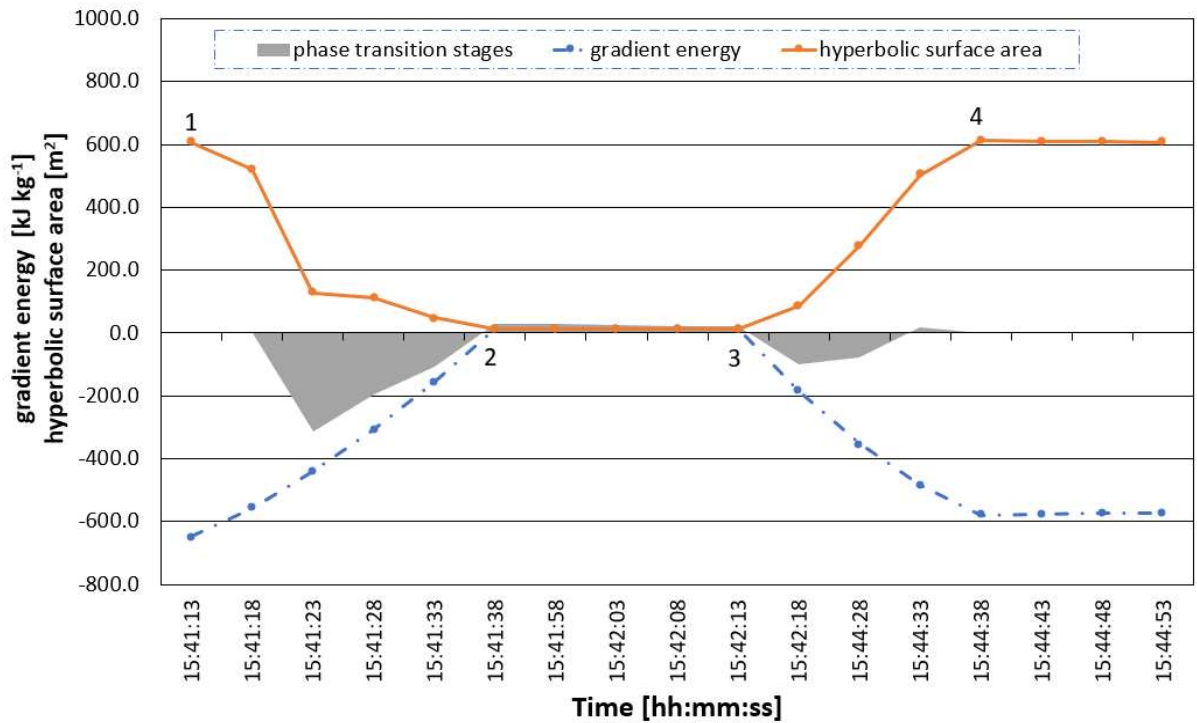

Fig. 8: Constrained mapping of gradient energy to hyperbolic surface area reveals phase transitions

If topological defects are admitted, then it becomes possible to remove the constraint $0 < r_2/r_1 < 1$ such that ($r_2 > r_1$) is admissible during phase transitions and the gradient energy maps almost exactly to the negative inverse of hyperbolic surface area, as **Fig. 9**:

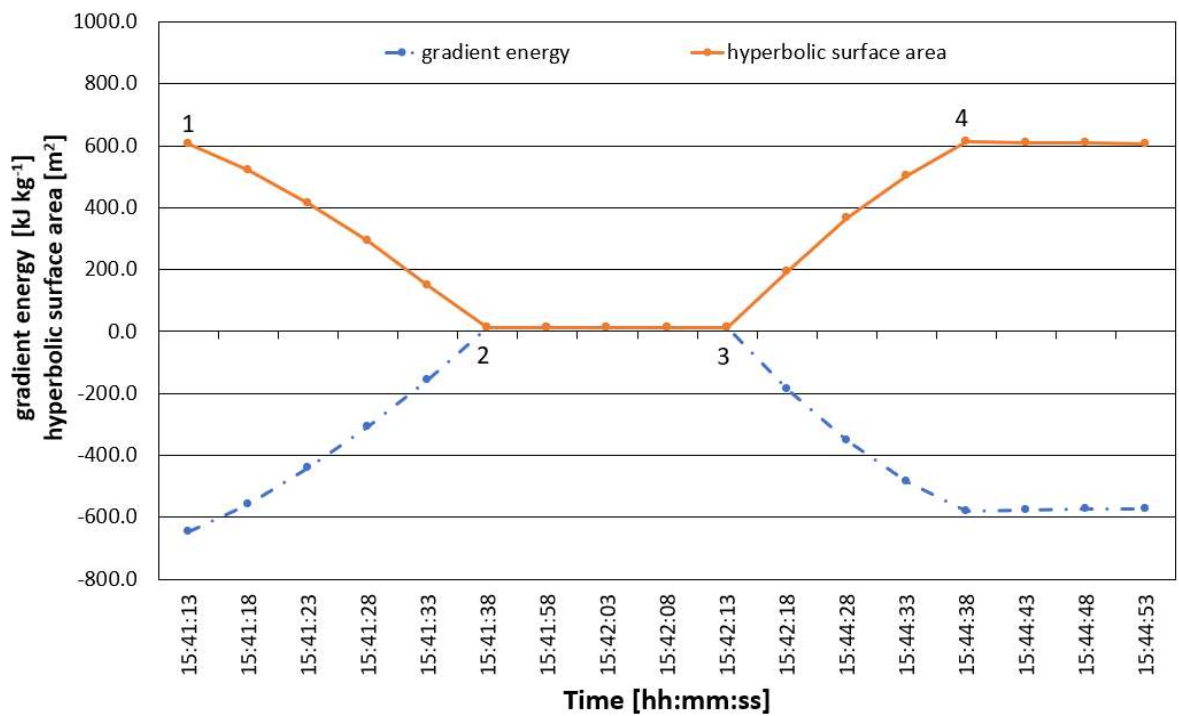

Fig. 9: Mapping of gradient energy to hyperbolic surface area with topological defects

**Fig. 10** displays the critical behaviour of reduced volume $|(V - V_c) / V_c|$ for the piston expander position with respect to the difference between the actual fluid temperature and critical temperature $|T - T_c|$, as determined by the critical exponents (**Table A4**). The resulting functional exponent of 1.779 is equal to the average correlation length exponent $v$ in 3-dimensions, ie. 3 x 0.593.

$V$ is the swept volume, $V_c$ is the critical volume, $T$ is the system temperature and $T_c$ is the critical temperature:

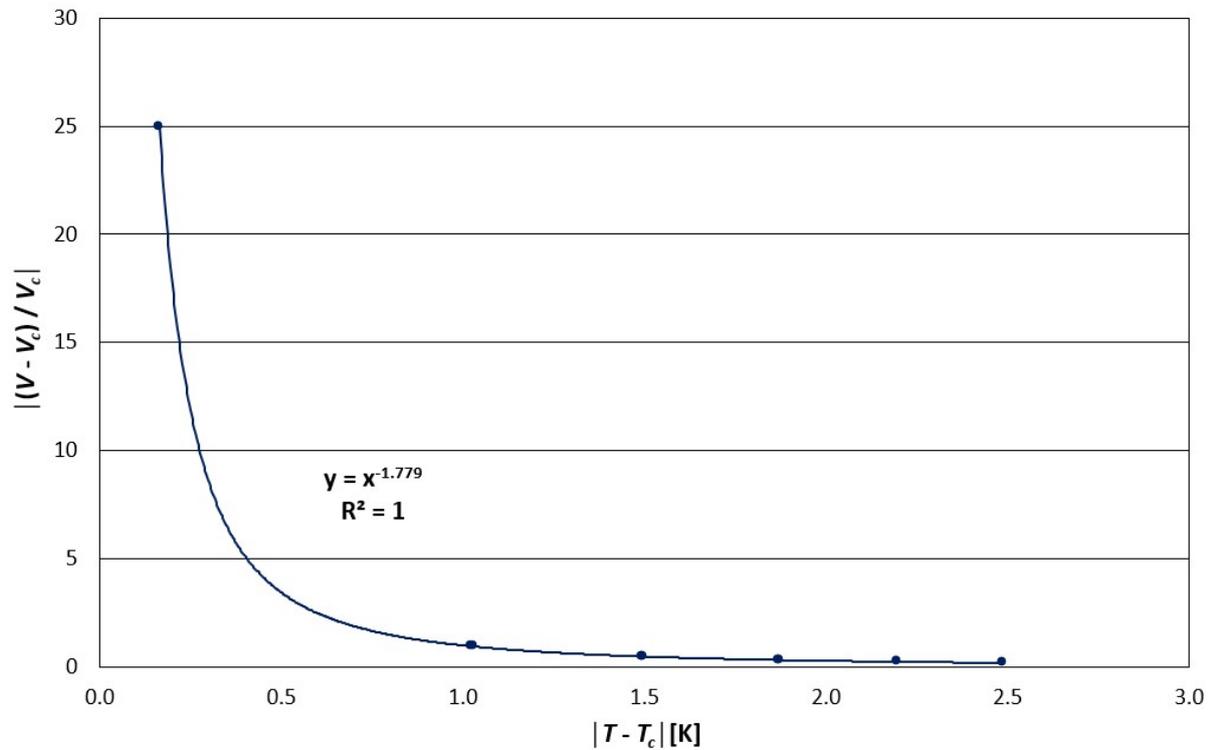

Fig. 10: Reduced volume as a function of temperature reveals the correlation length exponent

Variable susceptibility is indicative of phase transitions even though the fluid remains in a liquid phase with low variance in density. The hysteretic curves of swept volume with respect to fluid temperature in **Fig. 11** are also consistent with the presence of such liquid-liquid phase transitions. Since entropy and density display linear relationships with respect to temperature, changes in gradient suggest continuous, or second-order, phase transitions. Rest stages also indicate dielectric relaxation:

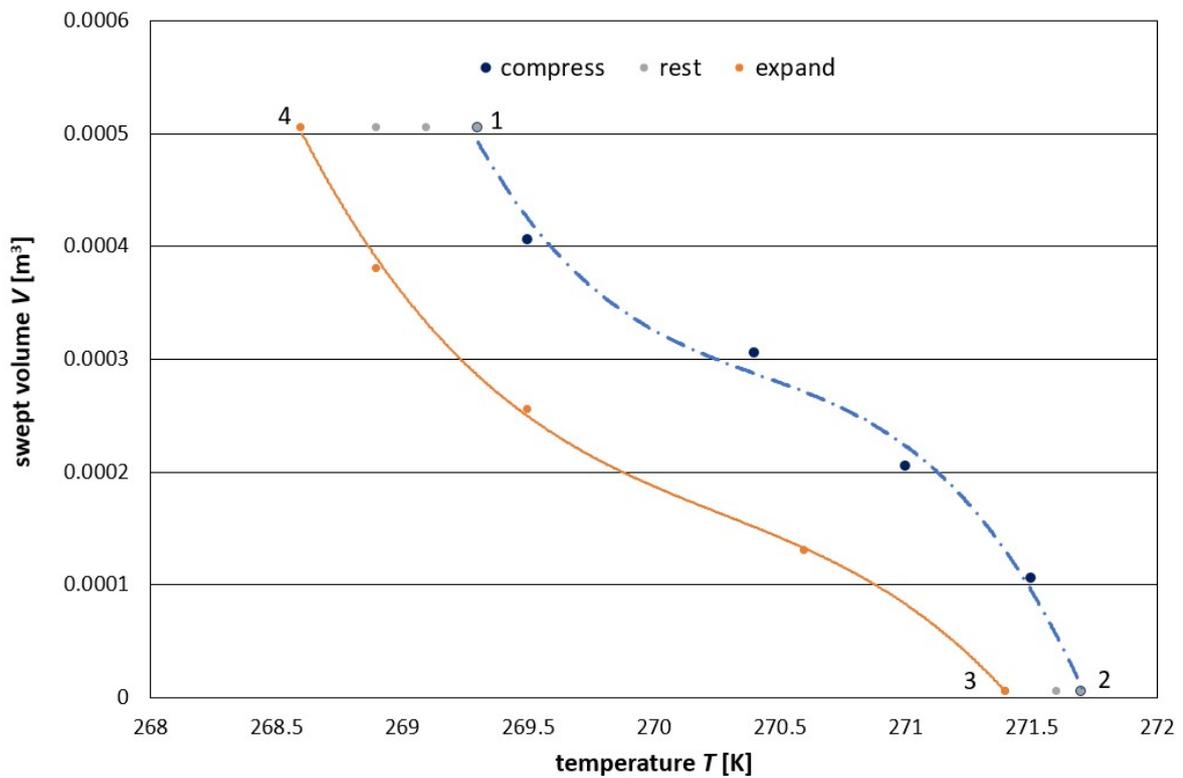

Fig. 11: Hysteresis loop suggesting continuous phase transitions and dielectric relaxation

A number of critical exponents can be derived for the phase-change behaviour indicated above from the REFPROP calculations (**Table A4**). The critical exponents allow the magnetic properties of the crystal-fluid to be determined. The plot associated with negative pressure perturbation resulting in positive piston displacement is shown in **Fig. 12**:

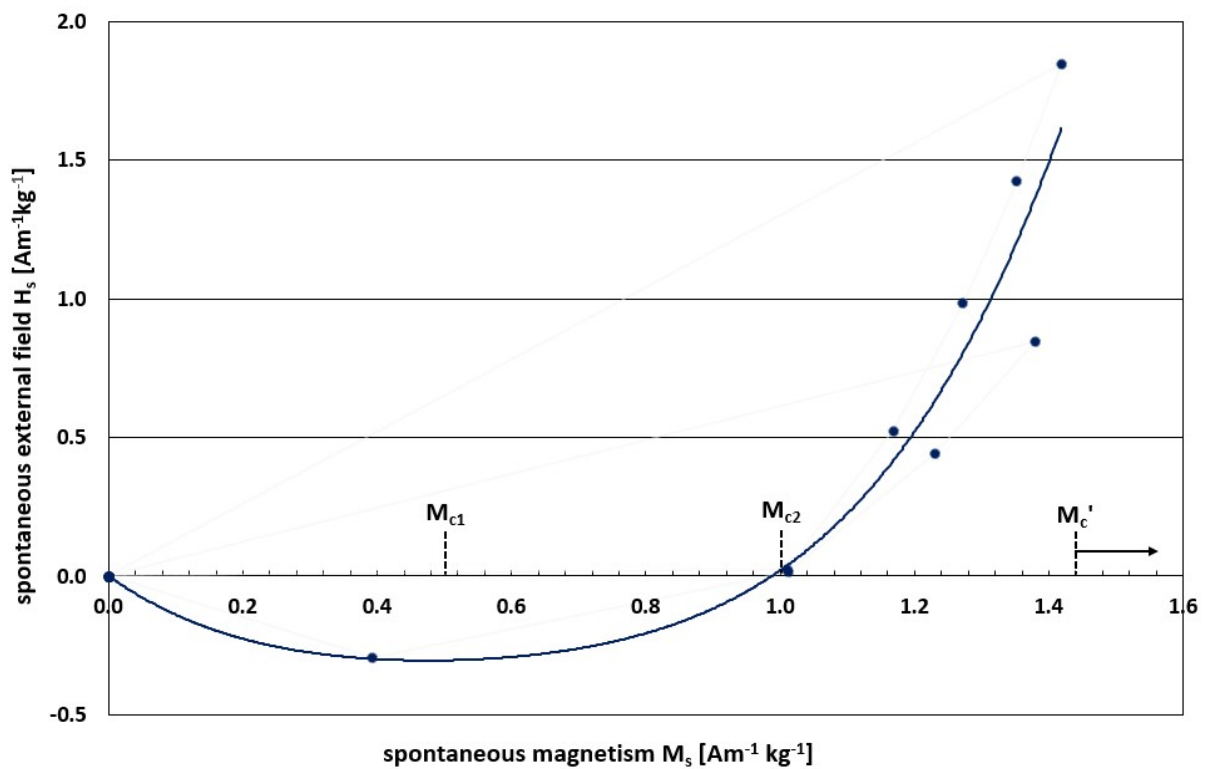

Fig. 12: Critical exponents reveal a magnetic phase transition and superconductor-like behaviour

The initial response between $0 < M_s < 0.5$ appears as diamagnetism ($H_s < 0$) and peaks at a lower critical field point $M_{c1}$, a result of high susceptibility $\chi$ such that $M_s \gg B_s$. Diamagnetism is then completely destroyed at $M_s = 1$, which represents an upper critical field point $M_{c2}$. For $H_s > 0$, a new value for the critical field $M_c'$ is deemed to be established to the right of the plotted values, the result of a phase transition at $H_s = 0$.

Analysis and discussion

A theoretical underpinning for the unusual experimental results is developed which seeks to establish a thermodynamic basis for the emergence of spontaneous magnetism and associated superconducting-like behaviours. Symmetry-breaking phase transitions then correspond to energy flows derived from a scalar field potential, ie. the embedding space of the crystal-fluid system. A background discussion relating to the non-equilibrium thermodynamics of inhomogeneities and constraints is given in **Appendix C** whilst the role of dissipative, supramolecular, clathrate cage structures is summarized in **Appendix D**.

Mathematical derivations for the critical exponent and correlation length scale values, together with the associated hyperbolic geometry can be found in **Appendices A**, **B** and **E**. These values derive from recorded experimental parameters and reveal strong correlations with the predictions of existing physical laws and theories such that no new theories need to be proposed. However, interesting insights into the effects of variable system inertia *I* (acceleration and deceleration) and correlation length scale $\xi$ on the magnitude of the scalar field are revealed, although the implications are not explored in the current work.

Both the quantitative and qualitative results summarized within the appendices signify the emergence of spontaneous magnetism and symmetry-breaking phase transitions in superconducting states. The theoretically predicted emergence of an associated gauge field able to couple with system electromagnetism represents a credible mechanism for harnessing a scalar field potential in the production of mechanical work. Thus, strong evidence exists that access to an unlimited and universal source of emissions-free, renewable energy has been achieved.

*Magnetism and critical phenomena*

With the pressure-perturbed establishment of magnetic exchange pathways, it is necessary to incorporate a magnetic term into the fundamental thermodynamic relation:

$$U = U(S, V, M, N) \tag{1}$$

where M is the extensive magnetic moment parameter. The intensive parameter conjugate to M is H, which generally is the external magnetic field that would exist in the absence of the system [42], which has zero value for the system under investigation. The relevant partial derivative is:

$$H = \left(\frac{\delta U}{\delta M}\right)_{S,V,N} \tag{2}$$

which, in the absence of thermodynamic constraints, would be expected to restore extensivity and additivity where the vdW interaction $F_{vdW}$ tends to zero such that:

$$\lim_{F_{vdW} \to 0} dU = TdS - Pdv + HdM + \sum_i \mu_i \, dn_i \tag{3}$$

However, recalling the affine entropy function in **Fig. 4** it is apparent that local stability conditions establish inhomogeneities that act to constrain both *v* and *u*. The flow of vacuum energy through the system as a response to critical behaviour is also deemed responsible for non-additivity in the corresponding Euler relation such that:

$$U \neq TS - PV + HM + \mu N$$

and for the Gibbs-Duhem relation:

$$S\mathrm{d}T - V\mathrm{d}P + \mathrm{M}\mathrm{d}H + N\mathrm{d}\mu \neq 0 \tag{5}$$

Whilst additivity would be restored where the flow of vacuum energy ceases (excluding any host-guest vdW interaction), an exchange mechanism is required to restore extensivity when the system constraint is relaxed or removed. That is, a mechanism is required to scale the interaction potential $\mu$. Before this can be quantitatively addressed, however, it will be necessary to evaluate the critical behaviour of the system.

The thermodynamic response functions in **Table A4** are calculated by the REFPROP program. These response functions enable a family of critical exponents to be derived [15,24]. The resulting critical exponents in **Table A4** fully satisfy the scaling laws of Fisher, Rushbrooke, Widom and Josephson, to produce a distinctive universality class in 3-dimensional space [15,24].

**Fig. 10** displays the critical behaviour of reduced volume $|(V - V_c) / V_c|$ for the piston expander position with respect to the difference between the actual fluid temperature and critical temperature $|T - T_c|$, as determined by the critical exponents. The resulting functional exponent of 1.779 is equal to the average correlation length exponent in 3-dimensions, ie. 3 x 0.593. Large spatial correlations are seen as $T_c \rightarrow T$ which represent the increasing length scale $\xi$ over which magnetic spins become correlated. The divergence in correlation length represented by $\xi$ reflects the presence of correlated spin patches of increasingly large size [25].

The critical system is highly sensitive to external pressure perturbations that appear to influence the extent of hydrogen bonded pathways, and thereby the flow of magnetic energy and entropy (ie. the interaction potential $\mu$, such that the constraint imposed on $v$ and $u$ appears tuneable). The varying magnetic susceptibility $\chi$ corresponds to large changes in spin correlations that produce divergence as $T_c \rightarrow T$. This manifests in either negative or positive piston displacements dependent upon the polarity of spontaneous magnetism $M_s$. However, it is apparent that $T_c$ is not a fixed value and that $T_c = T$ is found only at Points 1-4, the piston rest positions, and encompasses two stages of dielectric relaxation (2$\rightarrow$3 and 4$\rightarrow$1) [43,44].

During applied pressure perturbations, the system undergoes structural reorganizations that alter the value of $T_c$ such that it is continuously driven to attain stable states where $T_c = T$. The value of $T$ varies in response to variable external pressure so that $T_c$ lags $T$ in both directions to produce extended ranges of 'rolling' critical response [45] where the instantaneous magnitude of the response is determined by the dynamical value of $|T - T_c|$.

Generally, for systems in equilibrium, the critical point is reached by precisely tuning a control parameter [46]. However, for non-equilibrium systems, the critical point can be an attractor of the dynamics whilst remaining fully consistent with system parameters [47]. At any point where $T_c = T$ the system is in a stable, non-equilibrium state but remains on the threshold of instability. Small perturbations elicit critical behaviour where long-range interactions come to dominate the system, ie. the scale-invariant behaviour does not need to be precisely tuned to a critical value. This phenomenon is known as 'self-organized criticality'. Such systems are dynamically attracted to a state where they display power-law scaling, ie. long-range correlations bring about a new, effective interaction and global features that are very different from the underlying microscopic physics. It is characteristic for these systems to be slowly driven to an instability threshold with the subsequent energy-releasing reconfiguration occurring much more rapidly, ie. slow driving and fast relaxation [48].

**Appendix E** calculates the correlation length at 3.05. This equates to approximately 3.5 equal steps of 'rolling' critical response. The response functions and critical exponents of **Table A4** allow the magnetic properties of the system to be determined, as recorded in **Table A5**. The following Maxwell relations then provide for an exchange mechanism to equilibrate frustrated magnetic energy:

$$\left(\frac{\delta V}{\delta M}\right)_{S,P} = \left(\frac{\delta H}{\delta P}\right)_{S,M} \quad \text{and} \quad \left(\frac{\delta M}{\delta P}\right)_{S,H} = \left(\frac{\delta V}{\delta H}\right)_{S,P}$$

(6)

also

$$\left(\frac{\delta N}{\delta M}\right)_{S,\mu} = -\left(\frac{\delta H}{\delta \mu}\right)_{S,M} \quad \text{and} \quad \left(\frac{\delta M}{\delta \mu}\right)_{S,H} = -\left(\frac{\delta N}{\delta H}\right)_{S,\mu}$$

(7)

The results in **Table A6** reveal that *PV* work is only partially accounted for by relaxation of the constraint on *v* and *u*, as imposed by inhomogeneities and magnetic frustration. However, calculations based upon extensive parameters and intensive parameters only relate to the material system and do not account for the vacuum occupied by that system [42].

The total energy within the spatial region occupied by the system needs to include the vacuum energy value given by:

$$U_{vac} = \mu_0^{-1} H^2 V$$

(8)

Where $\mu_0$ is permeability of *flat*, free space with an historical value of $4\pi \times 10^{-7}$ T-m A$^{-1}$ [42]. However, as detailed in **Table A7**, the hyperbolic geometry of the system allows for effective values of $\mu_0$ which in fact reflect the ratio of Euclidean space to hyperbolic space. From the geometry of the system, the 'space density' is then found to be dependent upon the correlation length scale $\xi$ of 3.05, applicable to approximately 3.5 steps of 'rolling' critical response, such that;

$$e^{3.5\xi} = \mu_{0f}/\mu_{0i}$$

(9)

where $\mu_{0f}$ is the final effective value of $\mu_0$ and $\mu_{0i}$ is the initial effective value of $\mu_0$.

To establish vacuum energy as the conserved energy input that couples to the constant Hamiltonian oscillator, it is necessary to find a relationship between *PV* work, the relevant critical exponent, and the effective value of $\mu_0$. This is expounded in **Appendix B** where a gradient energy term, as derived from the Lagrangian function, describes the dynamical system. This, in turn, reveals scalar field $\Phi$ values associated with the critical response (**Table A8**) and a field coupling constant of 1.

Energy conservation requires that *PV* work be balanced through an equal-and-opposite electromagnetic coupling to the scalar field. The result obtained is in agreement with the cosmological model proposed by Ratra [49], ie:

$$\int P dV \propto e^{3v\Phi}$$

(10)

Values for Points 1-4 are given in **Table A8**. The gradient energy term of the Lagrangian also corresponds to the hyperbolic geometry and topology of the system, as well as its inertia. Details are given in **Appendix B** where gradient energy can be expressed on the surface area of a hyperbolic, hollow sphere. When considering the variable effective radius *r* of the system (ie. its inertia), the corresponding Euclidean volume ~ $e^{3r}$. The factor for Euclidean volume can then applied to establish the factor for effective vacuum permeability change:

for negative displacements $\quad \mu_{0f}/\mu_{0i} = e^{3r_i}/e^{3r_f}$

for positive displacements $\quad \mu_{0f}/\mu_{0i} = e^{3r_f}/e^{3r_i}$

(11)

When combining Einstein's mass-energy equivalence formula with Maxwell's electromagnetic wave equation, the significance of a constant Hamiltonian for the quasi-micro-canonical ensemble (as defined in **Appendix D**) becomes apparent (since total energy $E_1 = E$):

$$E_1 = \frac{E}{\sqrt{\left(\frac{\varepsilon_0}{\varepsilon_s}\right) \cdot \left(\frac{\mu_0}{\mu_s}\right)}}$$

(12)

The dynamics of the system stem from an exchange of the magnetic and electric components of electromagnetic energy. With external pressure perturbations, the magnetic flux increases through the spin correlations/ exchange interactions associated with the effective increase in $\mu_0$. Correspondingly, the electric flux reduces in response to an effective decrease in $\varepsilon_0$.

The quasi-micro-canonical ensemble can also be expressed in terms of an energy exchange between the magnetic B and electric E field components of the Lorentz invariant electromagnetic field pseudo-scalar [50]:

$$\frac{1}{2} F^{\mu\nu}F_{\mu\nu} = B^2 - \frac{E^2}{c^2} = \text{constant}$$

(13)

where *c* is the speed of light.

The covariant vector $F_{\mu\nu}$ and contravariant gradient potential $F^{\mu\nu}$ combine to produce Lorentz invariance for the pseudo-scalar field when rotated on a hyperbolic manifold, ie. the electromagnetic field pseudo-scalar corresponds to the geometry resulting from the non-extensivity and non-additivity of the non-equilibrium system. To account for the *PV* work extracted from the system with no apparent penalty on electromagnetic energy, the pseudo-scalar needs to be gauge-invariantly coupled to the scalar field described above, ie:

$$\int PdV = \frac{1}{2} F^{\mu\nu}F_{\mu\nu} e^{3v\Phi}$$

(14)

(where *V* is the specific swept volume )

and still corresponding to the cosmological inflation model proposed by Ratra [49].

These are transient effects that manifest only during external pressure perturbations. A positive pressure perturbation and converging magnetic flux are associated with negative work, whilst a negative pressure perturbation and diverging magnetic flux are associated with positive work. The effects in terms of the swept volume of the piston expander are shown in **Fig. 11**.

This coupling model suggests an explanation for the temperature and pressure changes recorded by the instrumentation associated with the sample vessel despite its mechanical isolation from the dynamics of the piston expander. The physical properties of the crystal-fluid appear not to be directly determined by the fluid dynamics of the system but rather mediated by coupling of the electromagnetic field pseudo-scalar to the scalar field.

### Effective magnetic monopoles and superconductivity

It appears that establishing integrated pathways for the energy flows associated with either negative or positive work distorts the tetrahedral lattice framework of the crystal-fluid. Under these conditions, the dangling hydrogen bonds of the excited state (or sterically restricted) clusters could overcome the local stability of geometrical frustration to form magnetic exchange pathways whilst simultaneously imposing hyperbolic geometry on the lattice manifold, as suggested in **Fig. 15** and **Fig. 16** (**Appendix D**).

**Fig. 12** reveals the relationship between pressure-induced spontaneous magnetism $M_s$ and the resulting magnetic field $H_s$. In this example, a negative pressure perturbation produces positive values of $M_s$. The initial response between $0 < M_s < 0.5$ is diamagnetic ($H_s < 0$), a result of high susceptibility $\chi$ such that $M_s \gg B_s$. The initial emergence of this spontaneous magnetism is attributed to the anisotropy of the clathrate hydrate structures and consistent with proton delocalization [35]. A similar situation is described by Purcell and Pound [51] where opposition of the external field to the magnetizing flux is attributed to the relaxation time for mutual interaction among spins being less than that between spins and the embedding lattice. This would be consistent with an increase in the spin correlation length $\xi$.

For $H_s < 0$, the profile resembles Type II superconductor behaviour [15], although it is large, rather than negative, values of susceptibility $\chi$ that are held responsible for the diamagnetic effect. The diamagnetism peaks at $M_{c1}$, the lower critical field, corresponding to minimal magnetic flux density $B_s$. Moving beyond $M_{c1}$ towards $M_{c2}$ the upper critical field, $B_s$ increases until all superconducting behaviour is destroyed at $M_s = 1$, $H_s = 0$. For $0 < M_s < 1$, the Ginzburg-Landau [52] parameter $\kappa$ is determined:

$$\kappa = \frac{M_{c2}}{\sqrt{2}M_{c1}} \text{ to give } \kappa > \frac{1}{\sqrt{2}}$$

(15)

ie. a Type II superconductor classification.

Taking $H_s$ as the order parameter, a phase transition is apparent at $M_{c2}$ where $H_s = 0$. For $M_s > 1$, $H_s$ takes positive values as the susceptibility $\chi$ falls below unity. A new value for the critical field $M_c'$ is deemed to be established to the right of the plotted values. It then follows that for this region $\kappa < 1/\sqrt{2}$ which is consistent with Type I superconducting behaviour. Here, large values of $B_s$ act to exclude the electric field E to establish dual superconducting behaviour [53,54].

Within the Type II region, susceptibility $\chi$ and the ratio of $H_s/M_s$ are seen to be greater than unity (ie. larger than ideal superconductor values). This result supports the inequality seen in the magnetic Maxwell relations (6,7) and provides further evidence for a vacuum energy contribution to the spontaneous magnetic flux $M_s$.

Progressive destruction of Type II superconductivity between $M_{c1}$ and $M_{c2}$ is consistent with the penetration of the magnetic field by Abrikosov vortices [15]. Above $M_{c2}$ it appears that the spins of these isolated vortices become aligned, ie. $M_{c2}$ corresponds to the correlation length $\xi$. Under these conditions, the current loops of the vortices are opposed and thus nullified inside the bulk material. However, surface loops do not cancel and the surface current generates a solenoidal effect that leads to an additional magnetic field [55].

Relatively recent advancements in the understanding of spin ice materials provide for an understanding of the superconducting phase transition described above, since effective monopole defects act as both divergent sources and convergent sinks of the magnetic field H [56]. Local proton ordering in water ice finds a physical equivalent in the frustrated pyrochlore Ising magnet, as originally identified by Anderson [13,14,57,58]. The Ising spin, as the proton spin, possesses only two discrete orientations (up or down) since it is constrained to point along an axis. The centres of the pyrochlore tetrahedra of these spin ice materials identify with the locations of oxygen atoms in water ice. The Ising spin axes are exactly aligned to the water ice oxygen bonding such that residual Pauling entropy can be precisely demonstrated in spin ices [59].

The high level of degeneracy expected for the tetrahedrally coordinated crystal-lattice being investigated (**Fig. 15** and **Fig. 16**) could lead to similar fractionalization of the magnetic moments into effective, irrationally-charged monopole excitations, ie. fractionalization of the microscopic spin degrees of freedom [60]. This behaviour is widely reported in low-energy, spin ice magnets [61,62, and references therein]. Crucially, gauge fields also emerge from the geometrical frustration and defects in these fields lead to further monopole excitations [60].

During the Type II superconducting phase described above, the order parameter is taken to be $H_s$ which goes to zero at $M_{c2}$ and so represents the point at which a gauge field should emerge. The reversal of $H_s$ at $M_{c2}$, from negative to positive, is accompanied by an apparent reversal in the order parameter. That is, a continuous phase transition with a corresponding broken symmetry:

$$H = \begin{cases} < 0, & M < 1, \quad M > -1 \\ > 0, & M > 1, \quad M < -1 \end{cases}$$

(16)

However, for a frustrated system with charge fractionalization, ordering is generally characterized by the emergent gauge field; ie. topological ordering [60,63], or a constrained disorder [59] that opposes ordering and symmetry breaking. Typically, in an ensemble subjected to constrained disorder, violations of local rules (or topological defects) appear as localized excitations of the low energy states of the system and these local rule violations control the collective dynamics. In spin ices and other water ice manifolds, the violations manifest as effective magnetic monopoles. Castelnovo *et al* [60] describe a mechanism for this fractionalizing process. The monopoles cut off the dipolar correlations to allow a correlation length $\xi$ to be defined. As a spin ice approaches a critical point, the spin correlation is not defined by an order parameter. Instead, it is defined by deconfinement of a gauge field which, in this case, is imposed by the quasi-micro-canonical ensemble/ constant Hamiltonian function. Such an interpretation is consistent with the convention that superconducting states lack any local order parameter but rather exhibit charge fractionalization whilst remaining sensitive to the global topology of the underlying manifold [64,65].

In many circumstances the effective monopoles of spin ice and water ice exhibit properties expected for the elusive, fundamental Dirac monopoles [66]. They exhibit magnetic charge and may be manipulated through means consistent with existence of a particle nature. In many situations, effective monopoles display behaviour consistent with the possession of inertia and mass [67].

Steinhardt proposes a relevant model for fundamental monopole (or vortex) dissociation and decay of a metastable false vacuum [68]. This model proposes that monopole cores act as inhomogeneous nucleation sites able to expand and convert the volume of the system to the phase represented by the core of the monopole. Several conditions are imposed so that the transition from the metastable phase to the stable phase can be mediated by this monopole dissociation:

1. There exists an isovector scalar field *Φ* and a scalar potential *V(Φ)*. The metastable phase corresponds to the local minima of *V(Φ)* which spontaneously breaks a continuous symmetry. The true phase will correspond to a global minimum of *V(Φ)*.
2. The monopole corresponds to a solution where the scalar field interpolates from a value near to the global minimum of *V(Φ)* at the core to a value equal to a local minimum of *V(Φ)* at distances from the core.
3. Inhomogeneous nucleation via monopoles should produce a finite density of topological excitations in the metastable system.

For the case being considered, effective monopole vortices would act to exclude the electric field to establish dual Type I superconductivity, in contrast to Abrikosov vortices which act to destroy Type II superconductivity [15]. The presence of a gradient energy term in the Lagrangian function (**Appendix B** and (17)) requires a scalar potential *V(Φ)* able to describe the thermodynamic response of the topologically ordered system displaying critical behaviour. It is the hyperbolic geometry of the system that determines the global minimum of $V(\Phi)$, ie. the value is determined by the effective value of $\mu_0$ which can be larger or smaller than the local minimum of $V(\Phi)$. Once electromagnetic coupling is established, energy can either flow into the monopoles to reduce system volume **V** or, alternatively, energy can flow out of the monopoles to increase system volume **V**.

Conclusion

A non-equilibrium, metastable system displaying non-extensivity and non-additivity is described. Local stability criteria are maintained through inhomogeneous combinations of system phases corresponding to a quasi-micro-canonical ensemble with constant Hamiltonian function.

Dissipative structures assembled from nanoscale clathrate hydrate cages, and having the potential to form magnetic exchange pathways, are integrated into a tetrahedrally-coordinated solvent inhibitor. This magnetically frustrated system displays self-organized criticality so that even when stable it remains on the threshold of instability. External pressure perturbations produce critical behaviour that can be described by a distinctive universality class of critical exponents, revealing a correlation length for long-range ordering.

Non-extensivity and critical behaviour produce hyperbolic curvature that acts upon the embedding space to establish variable, effective magnetic permeability. The Lagrangian function for the system reveals a relationship for the coupling of associated vacuum energy to the gradient energy term, enabling work outputs to result from the critical divergence.

Spontaneous magnetism resulting from critical behaviour reveals superconducting-like properties. The initial diamagnetic, Type II superconducting state is attributed to the anisotropy of the clathrate hydrate cages and the difference in relaxation times between magnetic spins and that between magnetic spins and the embedding lattice structure. The order parameter is identified as the spontaneous magnetic field $H_s$, which goes to zero as diamagnetism is progressively destroyed by Abrikosov vortices. Above the upper critical field value, Type I dual superconducting behaviour is observed, indicating bulk exclusion of the electric field. Although the spontaneous magnetic field $H_s$ still appears to be the order parameter for the dual superconducting state, ordering is instead attributed to the emergence of a gauge field and topological defects arising from magnetic frustration and charge fractionalization.

The topological defects are identified as effective magnetic monopoles. Deconfinement of the monopoles is deemed responsible for bulk exclusion of the electric field and emergence of the dual superconductor behaviour. The cosmological models proposed by Ratra and Steinhardt accurately describe the resulting coupling of a scalar potential of vacuum energy to the oscillating electromagnetism of the condensed matter system. The effective magnetic monopoles, or vortices, act as inhomogeneous nucleation sites able to expand and convert the volume of the system to the phase represented by the core of the monopole thereby performing mechanical work. Thus, the experimental results suggest that access to an unlimited and universal source of emissions-free, renewable energy has been achieved.

Acknowledgements

Appendix A

The recorded data, REFPROP [40] calculated properties, and other derived properties are presented in the tables below:

Table A1: Recorded and calculated thermodynamic properties of a typical piston expander cycle

|  | Temperature $T$ (K) | Pressure $P$ (MPa) | specific volume $v$ (m$^3$ kg$^{-1}$) |
|---|---|---|---|
| Point 1 | 269.3 | 0.29 | 0.0015 |
| Point 2 | 271.7 | 0.62 | 0.0015 |
| Point 3 | 271.4 | 0.61 | 0.0015 |
| Point 4 | 268.6 | 0.25 | 0.0015 |

|  | internal energy $u$ (kJ kg$^{-1}$) | entropy $s$ (kJ kg$^{-1}$ K$^{-1}$) | Volume $V$ (m$^3$) |
|---|---|---|---|
| Point 1 | -209.3 | 0.79 | 0.000505 |
| Point 2 | -200.6 | 0.83 | 0.000005 |
| Point 3 | -201.7 | 0.82 | 0.000005 |
| Point 4 | -211.9 | 0.78 | 0.000505 |

| displacement | $\Delta Ts$ heat (kJ kg$^{-1}$) | $\Delta Pv$ work (kJ kg$^{-1}$) | $\Delta PV$ work (kJ kg$^{-1}$) |
|---|---|---|---|
| -ve 1-2 | 1.93 | 0.0015 | 642.9 |
| +ve 3-4 | -2.23 | -0.0015 | -568.6 |

|  | enthalpy $h$ (kJ kg$^{-1}$) | Gibbs free energy $G$ (kJ kg$^{-1}$) | Helmholz free energy $F$ (kJ kg$^{-1}$) |
|---|---|---|---|
| Point 1 | -208.9 | -422.5 | -423.0 |
| Point 2 | -199.6 | -424.0 | -424.9 |
| Point 3 | -200.7 | -423.8 | -424.7 |
| Point 4 | -211.5 | -422.1 | -422.4 |

|  | 1 | 2 | 1 – 2 |
|---|---|---|---|
|  | $Pv$ from TD potentials (kJ kg$^{-1}$) | $Pv$ from REFPROP (kJ kg$^{-1}$) | Excess $Pv$ (kJ kg$^{-1}$) |
| Point 1 | 0.43 | 0.44 | -0.01 |
| Point 2 | 0.94 | 0.93 | 0.01 |
| Point 3 | 0.92 | 0.92 | 0 |
| Point 4 | 0.37 | 0.38 | -0.01 |

|  | 1 | 2 | 1 – 2 |
|---|---|---|---|
|  | $Ts$ from TD potentials (kJ kg$^{-1}$) | $Ts$ from REFPROP (kJ kg$^{-1}$) | excess $Ts$ (kJ kg$^{-1}$) |
| Point 1 | -422.1 | 213.7 | -635.8 |
| Point 2 | -423.0 | 224.4 | -647.4 |
| Point 3 | -422.8 | 223.0 | -645.8 |
| Point 4 | -421.7 | 210.5 | -632.2 |

|         | 1                        | 2                        | 3                  | 1 − 2 + 3                |
|---------|--------------------------|--------------------------|--------------------|--------------------------|
|         | excess $Ts$<br>(kJ kg$^{-1}$) | excess $Pv$<br>(kJ kg$^{-1}$) | $G$<br>(kJ kg$^{-1}$) | excess $u$<br>(kJ kg$^{-1}$) |
| Point 1 | -635.8                   | -0.01                    | -422.5             | -1058.3                  |
| Point 2 | -647.4                   | 0.01                     | -424.0             | -1071.4                  |
| Point 3 | -645.8                   | 0                        | -423.8             | -1069.6                  |
| Point 4 | -632.2                   | -0.01                    | -422.1             | -1054.3                  |

Table A2: Gradient energy where the Gibbs energy is associated with a phase-change process

|         | 1                                     | 2                               | 1 - 2                    |
|---------|---------------------------------------|---------------------------------|--------------------------|
|         | $u$ (from TD potentials) - $G$<br>(kJ kg$^{-1}$) | internal energy $u$<br>(kJ kg$^{-1}$) | Excess $u$<br>(kJ kg$^{-1}$) |
| Point 1 | -422.6                                | -209.3                          | -213.3                   |
| Point 2 | -424.0                                | -200.6                          | -223.4                   |
| Point 3 | -423.8                                | -201.7                          | -222.1                   |
| Point 4 | -422.0                                | -211.9                          | -210.1                   |

|         | 1                              | 2                        | 3                       | 1 - 2 - 3                                   |
|---------|--------------------------------|--------------------------|-------------------------|---------------------------------------------|
|         | internal energy $u$<br>(kJ kg$^{-1}$) | excess $u$<br>(kJ kg$^{-1}$) | $\int PdV$<br>(kJ kg$^{-1}$) | gradient energy $^§$<br>½$(\nabla\Phi)^2$<br>(kJ kg$^{-1}$) |
| Point 1 | -209.3                         | -213.3                   | 651.8                   | -647.9                                      |
| Point 2 | -200.6                         | -223.4                   | 8.9                     | -5.0                                        |
| Point 3 | -201.7                         | -222.1                   | 8.3                     | -4.4                                        |
| Point 4 | -211.9                         | -210.1                   | 576.9                   | -578.7                                      |

$^§$ includes work associated with the walls of the vessel.

Table A3: Calculation of hyperbolic surface area

|         | 1                                         | 2                                  | 1 - 2                                  |                                                    |
|---------|-------------------------------------------|------------------------------------|----------------------------------------|----------------------------------------------------|
|         | outer radius<br>$r_1 = 5/(2Pv)$<br>(m kg$^{-1}$) | inner radius $r_2$<br>(m kg$^{-1}$) | effective radius<br>$R = r_1 - r_2$<br>(m kg$^{-1}$) | surface area<br>$4\pi\sinh^2(R/2)$<br>(m$^2$ kg$^{-1}$) |
| Point 1 | 5.8                                       | 0.5                                | 5.3                                    | 606.2                                              |
| Point 2 | 2.7                                       | 0.8                                | 1.8                                    | 14.0                                               |
| Point 3 | 2.7                                       | 1.0                                | 1.7                                    | 11.6                                               |
| Point 4 | 6.8                                       | 1.5                                | 5.3                                    | 611.9                                              |

Derivation of the effective system radii is detailed in **Appendix E**.

Table A4: Response functions and critical exponent family as calculated by REFPROP

|         | density<br>$\rho$ (kg m$^{-3}$) | specific heat capacity<br>$C_v$ (kJ kg$^{-1}$ K$^{-1}$) | specific heat capacity<br>$C_p$ (kJ kg$^{-1}$ K$^{-1}$) | isothermal compressibility<br>$K_T$ (kPa$^{-1}$) |
|---------|---------------------------------|---------------------------------------------------------|---------------------------------------------------------|-------------------------------------------------|
| Point 1 | 665.1                           | 2.871                                                   | 3.714                                                   | 0.4557                                          |
| Point 2 | 663.7                           | 2.850                                                   | 3.698                                                   | 0.4484                                          |
| Point 3 | 663.9                           | 2.844                                                   | 3.694                                                   | 0.4462                                          |
| Point 4 | 665.6                           | 2.869                                                   | 3.712                                                   | 0.4547                                          |

|  | Critical Exponent | | |
|---|---|---|---|
| displacement | heat capacity $\alpha$ ($C_v$ / $C_p$) | order parameter $\beta$ | susceptibility $\gamma$ |
| -ve 1-2 ($T - T_{crit}$) | 0.222 / 0.286 | 0.384 | 0.916 |
| +ve 3-4 ($T - T_{crit}$) | 0.223 / 0.287 | 0.515 | 0.765 |

|  | Critical Exponent | | |
|---|---|---|---|
| displacement | equation of state $\delta$ | correlation length $\nu$ ($C_v$ / $C_p$) | power law decay $\rho$ ($C_v$ / $C_p$) |
| -ve 1-2 ($T - T_{crit}$) | 3.385 | 0.593 / 0.571 | 0.455 / 0.396 |
| +ve 3-4 ($T - T_{crit}$) | 2.485 | 0.592 / 0.571 | 0.708 / 0.660 |

Corresponding scaling laws

$\gamma = \nu(2 - \rho)$     Fisher law

$\alpha + 2\beta + \gamma = 2$     Rushbrooke law

$\gamma = \beta(\delta - 1)$     Widom law

$\nu d = 2 - \alpha$     Josephson law

where d is the spatial dimension.

Table A5: Magnetic properties derived from critical exponents

|  | reduced volume $(V-V_c)/V_c$ | spontaneous magnetism $M_s$ (A m$^{-1}$ kg$^{-1}$) | spontaneous magnetic induction $B_s$ (T kg$^{-1}$) | spontaneous external field $H_s$ (A m$^{-1}$ kg$^{-1}$) |
|---|---|---|---|---|
| Point 1* | 0.198 | 1.418 | 3.265 | 1.846 |
| Point 2 | 0.952 | 1.011 | 1.036 | 0.026 |
| Point 3* | 25.0 | 0.394 | 0.099 | -0.295 |
| Point 4 | 0.329 | 1.380 | 1.380 | 0.845 |

* Initial value recorded 5s after pressure perturbation.

Table A6: Maxwell relations derived from Table A1 and Table A5

|  | 1 | 2 | |2| + |1| | |1| : |2| |
|---|---|---|---|---|
| displacement | $\left(\frac{\delta V}{\delta M}\right)_{S,P}$ (kJ kg$^{-1}$) | $\left(\frac{\delta H}{\delta P}\right)_{S,M}$ (kJ kg$^{-1}$) | inequality (kJ kg$^{-1}$) | ratio |
| -ve 1-2 | 1.23 x 10$^{-3}$ | -5.52 x 10$^{-3}$ | 6.75 x 10$^{-3}$ | 1 : 4.5 |
| +ve 3-4 | 5.07 x 10$^{-4}$ | -3.17 x 10$^{-3}$ | 3.68 x 10$^{-3}$ | 1 : 6.3 |

| displacement | 3 $\left(\dfrac{\delta M}{\delta P}\right)_{S,H}$ (kJ kg$^{-1}$) | 4 $\left(\dfrac{\delta V}{\delta H}\right)_{S,P}$ (kJ kg$^{-1}$) | \|3\| + \|4\| inequality $u_{vac}$ (kJ kg$^{-1}$) | \|3\| : \|4\| ratio |
|---|---|---|---|---|
| -ve 1-2 | -1.23 x 10$^{-3}$ | 2.75 x 10$^{-4}$ | 1.51 x 10$^{-3}$ | 4.5 : 1 |
| +ve 3-4 | -2.74 x 10$^{-3}$ | 4.39 x 10$^{-4}$ | 3.18 x 10$^{-3}$ | 6.3 : 1 |

The difference in the inequalities is attributed to timing inconsistencies when recording the magnetic properties noted in **Table A5**. Values for vacuum energy contributions would be based upon the following relationship:

$$U_{vac} = \frac{1}{2}\mu_0^{-1} H^2 V$$

(8)

Table A7: Effective Euclidean volume vs. hyperbolic volume

|  | net radius $r = r_1 - r_2$ | Euclidean volume ~ e$^{3r}$ | Euclidean volume factor ($f_1$) vs. min. at Point 3[†] |
|---|---|---|---|
| Point 1 | 5.3 | 7409663 | 44271 |
| Point 2 | 1.8 | 250 | 1.3 |
| Point 3 [†] | 1.7 | 167 | 1.0 |
| Point 4 | 5.3 | 7618473 | 45519 |

Vacuum permeability $\mu_0$ is, historically, a universal constant with a value of 4$\Pi$ x 10$^{-7}$ T-m A$^{-1}$ in flat, free space [42]. However, the increase in effective Euclidean volume directly relates to the increase in effective permeability, ie. the vacuum energy potential increases proportionally with effective Euclidean volume.

| displacement | space density factor ($f_1$) | ln ($f_1$) | ln ($f_1$)/ 3.5 SOC steps |
|---|---|---|---|
| -ve 1-2 | 44271 | 10.70 | 3.06 [‡] |
| +ve 3-4 | 45519 | 10.73 | 3.06 [‡] |

[‡] SOC – self-organized criticality – values consistent with the derived correlation length $\xi$ = 3.05, as detailed in **Appendix E**.

Table A8: Comparison of calculated gradient energy with calculated coupling energy

|  | gradient energy[^] ½($\nabla\Phi$)$^2$ [kJ kg$^{-1}$] | scalar potential[^] $\nabla\Phi$ [km s$^{-1}$] | scalar field $\Phi$ [ks$^{-1}$] | coupling energy e$^{3v\Phi}$ [kJ kg$^{-1}$] |
|---|---|---|---|---|
| Point 1 | -651.8 | -36.1 | 3.6 | 651.8 |
| Point 2 | -8.9 | -4.2 | 1.2 | 8.9 |
| Point 3 | -8.3 | 4.1 | 1.2 | 8.3 |
| Point 4 | -576.9 | 34.0 | 3.6 | 576.9 |

[^] excludes work associated with the walls of the vessel.

Appendix B

Since Steinhardt's model for monopoles [68] requires the existence of a scalar potential $V(\Phi)$, incorporation of a gradient energy term in the Lagrangian function describing the critical regions of the system should be consistent with the existence of such entities.

*PV* work is taken to conform to the principle of least action. Terms accounting for the kinetic, potential and interaction energies of the system are arranged to reveal a gradient energy term:

$$\frac{1}{2}(\nabla \Phi)^2 = u - u_e - \int P \mathrm{d}V \tag{17}$$

where $\frac{1}{2}(\nabla\Phi)^2$ is the gradient energy, $u$ is internal energy, and $u_e$ is excess internal energy. The calculated data are included in **Table A2**. Where work associated with the walls of the vessel is excluded, $u = u_e$, to give the simple relationship:

$$\frac{1}{2}(\nabla \Phi)^2 = -\int P \mathrm{d}V \tag{18}$$

The gradient energy is expressed in hyperbolic geometry, as revealed in **Fig. 9**, where dimensional analysis provides for a direct comparison. This is established through a physical response to interactions between excess negative energy potential, system inertia and surface area, as detailed below.

The excess negative potential manifests as non-extensivity in the fundamental thermodynamic relation, ie. the disparity between the specific volume and the specific swept volume of the system, such that the negative energy potential associated with non-extensivity is proportional to $Pv$. Since system inertia ($I = \frac{1}{2}mr^2$) is proportional to kinetic energy, which is in turn proportional to $1/Pv$, the average 1-dimensional radius $r_x$ of the stable, non-critical system is found:

$$r_x \propto \sqrt{\frac{2}{Pv}} \tag{19}$$

The Gaussian curvature $K$ for the 2-dimensional, hyperbolic surface of the non-critical system (ie. with no topological defects), for the principle curvature relationship of $r_x = -r_y$, is then determined:

$$K = \frac{1}{r_x r_y} \tag{20}$$

or

$$K \propto -\left(\frac{Pv}{2}\right) \tag{21}$$

Then, the average Gaussian radius of hyperbolic curvature ($1/K$ or $R_g$) is given by:

$$R_g = \left(\frac{2}{Pv}\right) \tag{22}$$

The critical behaviour acts to confine the electric field potential to the surface area of spin patches. Since the Gaussian radius of curvature $R_g$ effectively enables the hyperbolic surface area of a hollow, walled sphere having effective radius $R$ to be determined ($A = 4\pi\sinh^2(R/2)$), it can be compared to the gradient energy.

The outer radius $r_1$ is fixed as that of the solid sphere:

$$r_1 = \frac{5}{2Pv}$$

(23)

Values for the variable, inner radius $r_2$ are found iteratively such that $A \rightarrow \tfrac{1}{2}V(\Phi)^2$ as far as the constraint $0 < r_2/r_1 < 1$ permits. The calculated values associated with Points 1-4 are included in **Table A3**. The effective radius $R$ corresponds to variable inertia and negative curvature to describe a constant energy system that is decelerated under expansion (increasing inertia and increasing negative curvature) and accelerated under compression (decreasing inertia and decreasing negative curvature) with associated fictitious forces.

However, the emergence of effective magnetic monopoles should produce topological defects such that the geometrical constraint of $0 < r_2/r_1 < 1$ is overcome, ie. $r_2 > r_1$ is admissible, so that gradient energy can be fully expressed on the surface area of the hyperbolic manifold (see **Fig. 8** and **Fig. 9**).

For a 2-dimensional surface in 3-dimensional space (ie. a Type I superconductor), the Gaussian curvature $K$ is equal to the Ricci curvature $R_r$, ie. it is an invariant thermodynamic scalar function. For Ruppeiner and co-investigators [69-71], the thermodynamic curvature of a system is equal to its correlation volume near the critical point and, as for other critical phenomena, the Ricci scalar curvature $R_r$ diverges at a liquid-liquid critical point. $K$ thereby reveals information about intermolecular interactions and the size of organized, mesoscopic structures. It should also possible to detect a sign change for $K$; from negative on approaching the critical point, to positive on moving away from it. Under the Weinberg convention [72], a negative $R_r$ is associated with attractive interactions with a closed geometry (eg. a sphere) whilst a positive $R_r$ is associated with repulsive interactions and an open geometry (eg. a pseudo-sphere), as **Fig. 13**:

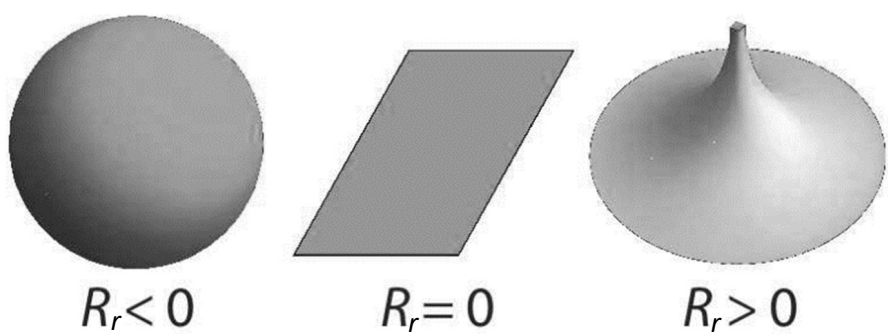

Fig. 13: The Ricci scalar curvature and corresponding geometry

Appendix C

*Inhomogeneities and constraints*

Callen [73] states that, 'the single all-encompassing problem of thermodynamics is the determination of the equilibrium state that eventually results after the removal of internal constraints in a closed, composite system'. The solution to this problem is then framed in terms of a number of entropy maximum postulates, which lead to the following equivalent extrema principles:

*Entropy Maximum Principle* – the equilibrium value of any unconstrained internal parameter is such as to maximize the entropy for the given value of the total internal energy.

*Energy Minimum Principle* – the equilibrium value of any unconstrained internal parameter is such as to minimize the energy for the given value of the total entropy.

The equivalence requires that the entropy of a composite system be additive over the constituent subsystems. The entropy must be continuous and differentiable, and a monotonically increasing function of the energy. The additivity property requires that the entropy of a simple system is a homogeneous first-order function of the extensive parameters [74] such that if all the extensive parameters of a system are multiplied by a constant λ, the entropy is multiplied by this same constant:

$$S(\lambda U, \lambda V, \lambda N_1, ..., \lambda N_r) = \lambda S(U, V, N_1, ..., N_r) \tag{24}$$

Additionally, the monotonic property as postulated implies:

$$\left(\frac{\delta S}{\delta U}\right)_{V_1, N_1, ..., N_r} > 0 \tag{25}$$

so that the condition of stability manifests in the concavity of the entropy function, ie. for two identical, adiabatic subsystems:

$$S(U + \Delta U, V, N) + S(U - \Delta U, V, N) \leq 2S(U, V, N) \tag{26}$$

Thus, it is apparent that the affine entropy function produced in **Fig. 4** masks an unstable, convex, fundamental relation [75]. The straight line represents local stability conditions achieved as a result of inhomogeneous combinations of at least two system phases. This loss of homogeneity is characteristic of phase transitions that give rise to critical phenomena. Local stability conditions ensure that inhomogeneities of either specific internal energy $u$ or specific volume $v$ separately do not increase entropy, and that a coupled inhomogeneity of $u$ and $v$ together does not increase overall entropy $s$. Specific volume $v$ and internal energy $u$ are thereby constrained, intensive parameters.

Other requirements for stability are the convexity of the thermodynamic potentials (internal energy $u$, enthalpy $h$, Helmholtz free energy $F$ and Gibbs free energy $G$) with respect to their extensive parameters, and concavity with respect to their intensive parameters. The impact of local stability conditions on thermodynamic potentials reached experimentally is presented in **Fig. 5** where the resulting excess internal energy potential $u_e$ mirrors the system's internal energy $u$, as calculated by REFPROP and recorded in **Table A1** and **Table A2**. For all the thermodynamic potentials, the experimental results produce affine functions such that extensive parameters are effectively indistinguishable from intensive parameters due to dynamically-responsive inhomogeneities.

Stability criteria are described qualitatively by Le Chatelier's Principle [75]. It states that any inhomogeneity developing in a system should induce a process that tends to eradicate that inhomogeneity. However, the flattening of the minimum of the Gibbs potential in the critical region of a phase transition implies the absence of a 'restoring force' for fluctuations emanating from the critical state [76]. This leads to divergent fluctuations in the relevant order parameters, such as the volume $V$ or the magnetic moment M. The dominant effect is the onset of long-range, correlated behaviour in the vicinity of the critical point. This is a consequence of long-wavelength interactions being the most easily excited. As fluctuations grow, the longest wavelength fluctuations grow fastest and come to dominate the properties of the critical region [76].

Such conditions might appear favourable for the growth of undistorted, long-range vdW interactions between guest-free clathrate hydrate structures and their former methane guest molecules, as previously reported [9,77]. However, the critical response observed in the latest experiments is transient and confined to a 'rolling' critical region. In fact, the calculated critical exponents and magnetic properties in **Table A4** (plotted in **Fig. 12**) reveal both diamagnetic-like and paramagnetic-like properties. Therefore, any long-range vdW $1/\sqrt{x}$ interaction is seen to be independent of any critical response and is in instead likely to be a consequence of the reduced electrostatic screening lengths associated with high ionic dissociation in the 'dielectric solvent' inhibitor [9,77]. Additionally, none of the critical exponents revealed displays an exact value of -½.

Appendix D

*Nanoscale clathrate hydrate cages*

Naively, the guest-free clathrate structures are initially formed as methane hydrates with 12 Å cubic crystal structure 8X · 46H$_2$O, or (CH$_4$)$_8$(H$_2$O)$_{46}$ [1,4] before the guest molecules are evacuated through exposure to low-energy, negative-pressure conditions. This basic packing structure is known as Structure I with a central (outlined) dodecahedron being surrounded by eight tetradecahedra [1,4].

The pentagonal dodecahedra are linked through 20 vertices by hydrogen bonds to form a pseudo-body-centred array in which central polyhedral are rotated 90° with respect to those at the origins. The surrounding tetradecahedra form columns along their faces which are arranged vertically, horizontally and out-of-plane. Each water molecule acts as acceptor and donor of hydrogen bonds (ie. every oxygen atom is four-fold coordinated) to produce a tetrahedral environment of hydrogen bonds within the hexagonal ice.

Zakharov [78], however, reports on experimental results obtained by x-ray scattering and spectroscopy techniques revealing that water on a nanometre scale is a fluctuating mixture of clusters with either tetrahedral structure or with partially broken, 'dangling', hydrogen bonds. The structural partition is attributed to the existence of the two distinct nuclear spin isomers; *ortho*-water and *para*-water [79], which do not interconvert in isolated molecules [80].

The two forms are distinguished by the quantum number values for total nuclear spin I, with I = 0 for *para*-water and I = 1 for *ortho*-water, ie. a magnetic moment arises only in the *ortho*-water isomer. The generalized Pauli principle requires that the total molecular wavefunction must be anti-symmetric under the permutation of the two hydrogen nuclei in a water molecule. *Para*-water is associated with symmetric rotational functions in the electronic-vibrational ground state of the molecule, whilst *ortho*-water is associated with anti-symmetric rotational functions. Nuclear spin and rotational symmetry are thus intimately linked to establish the following ground states [80,81]:

$$para - \text{water:} \quad |j_{K_a K_c}\rangle = |0_{00}\rangle \quad \text{(absolute ground state)}$$

$$ortho - \text{water:} \quad |j_{K_a K_c}\rangle = |1_{01}\rangle \quad \text{(first excited rotational state)}$$

(27)

Where $j$ is the quantum number of the rotational angular momentum and $K_a$ and $K_c$ are the quantum numbers for the projection of the rotational angular momentum on the principal axes of molecular inertia, *a* and *c*. The result is the long-range proton disorder originally identified by Pauling [11,82,83] that is responsible for residual entropy in water-ice and allows for disordered, effective ground states with macroscopic degeneracy, ie. a large number of states possessing the same energy.

Zakharov [78] then proposes that the absolute ground state is favourable for the formation of clusters with closed hydrogen bonds whilst the excited state gives rise to clusters with impaired bonding. These impaired, or dangling, hydrogen bonds [84], are O-H bonds that point towards the inhibitor solvent, rather than hydrogen bond to water. The hydrogen bonded lattice of the inhibitor solvent is also tetrahedrally coordinated such that it can readily incorporate the dangling hydrogen bonds to establish potential exchange pathways for the transfer of spin rotational energy to the environment [78]. Initially, however, the tetrahedral configuration serves to frustrate any spin alignment between the *ortho*-water clusters, ie. tetrahedral symmetry precludes the possibility that every pairwise interaction of proton spins can be satisfied at the same time [13].

The results in **Table A1, Table A2** and **Fig. 5** reveal the constant Hamiltonian function of the crystal-fluid lattice. It can be described as a quasi-micro-canonical ensemble in that $N$, $E$ and $v$ (rather than $V$) are constant when work associated with the walls of the vessel is excluded. As previously observed [9], a non-equilibrium system such as the one under investigation is not strictly isolated and additional sources of energy are required to establish energy conservation in cyclic processes. However, the current investigations provide further insight into these quasi-micro-canonical ensembles. It is proposed that there are three integrated but distinguishable sub-systems required for performance of the work cycle described in **Fig. 6** and **Fig. 11**:

> *A constant energy oscillator*, or quasi-micro-canonical ensemble, in which $v$ and $u$ are constrained and local stability conditions are maintained through inhomogeneous combinations of at least two system phases. When stable, the system remains on the threshold of instability such that external pressure perturbations produce divergent, critical behaviour enabling either negative or positive *PV* work.

> *An energy conserving process.* The *PV* work and energy conserving input associated with the critical response are considered to be an additional ensemble that becomes coupled to the quasi-micro-canonical ensemble. Non-extensivity, magnetic frustration and degeneracy produce large hyperbolic distortions in system geometry. These phenomena together with an emergent gauge field allow for electromagnetic coupling of the constant energy oscillator to the vacuum energy of the embedding space, ie. vacuum energy represents the conserved energy input.

> *A heat bath*, or quasi-canonical ensemble, where $N$, $T$ and $v$ are approximately constant, provides for a magnetocaloric exchange of entropy such that the heat bath acts as both ideal sink and ideal source of entropy. The sink and source are essential to the performance of the irreversible work cycle in accordance with the second law of thermodynamics.

The configuration of these three sub-systems is suggested schematically in **Fig. 14** below:

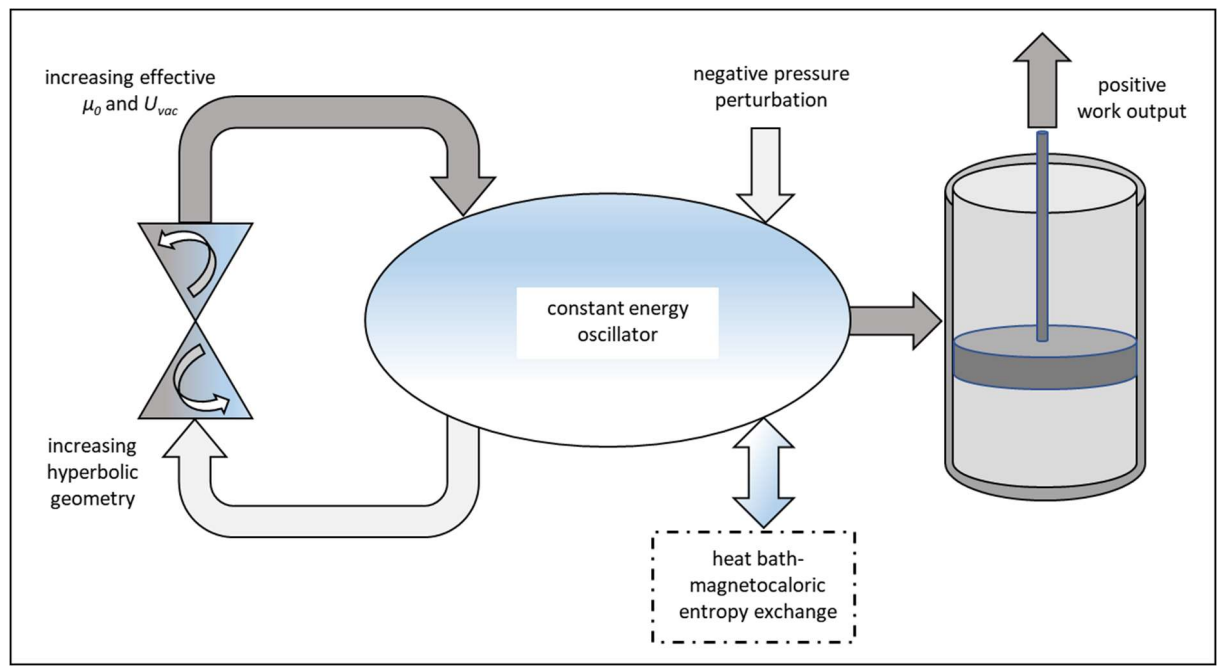

Fig. 14: Interaction of work cycle sub-systems for the case of positive work

Calculated results in **Table A3** reveal that the lattice distortions produce hyperbolic, or negative, curvature. **Fig. 15** and **Fig. 16** below provide schematic interpretations:

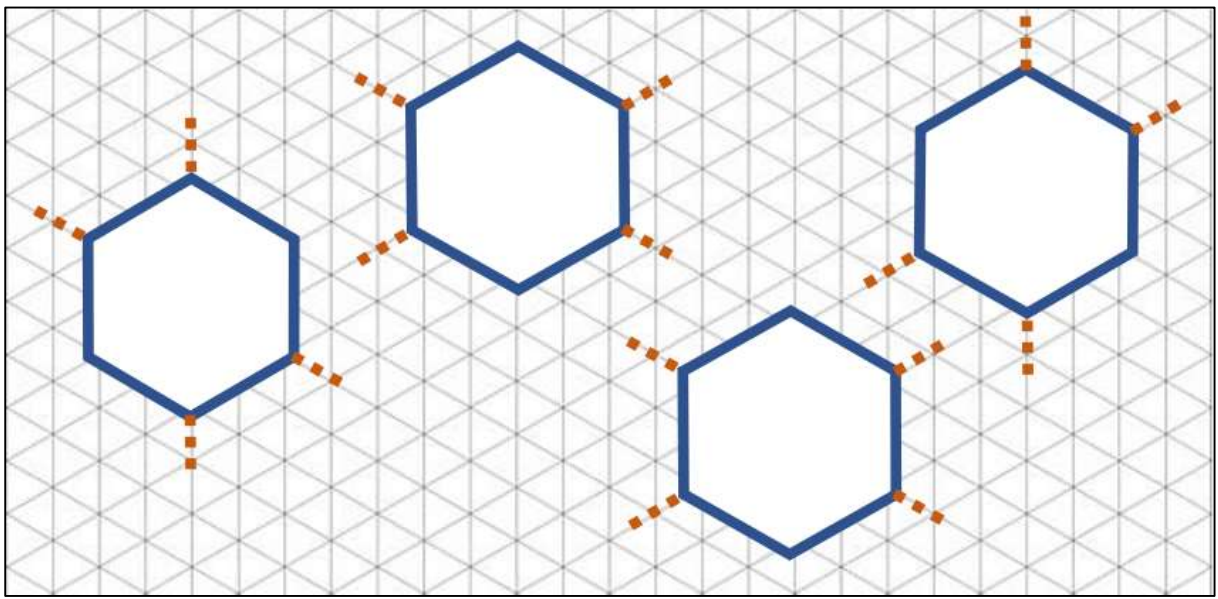

Fig. 15: Isolated clathrate hydrates with dangling H-bonds (dashed lines)

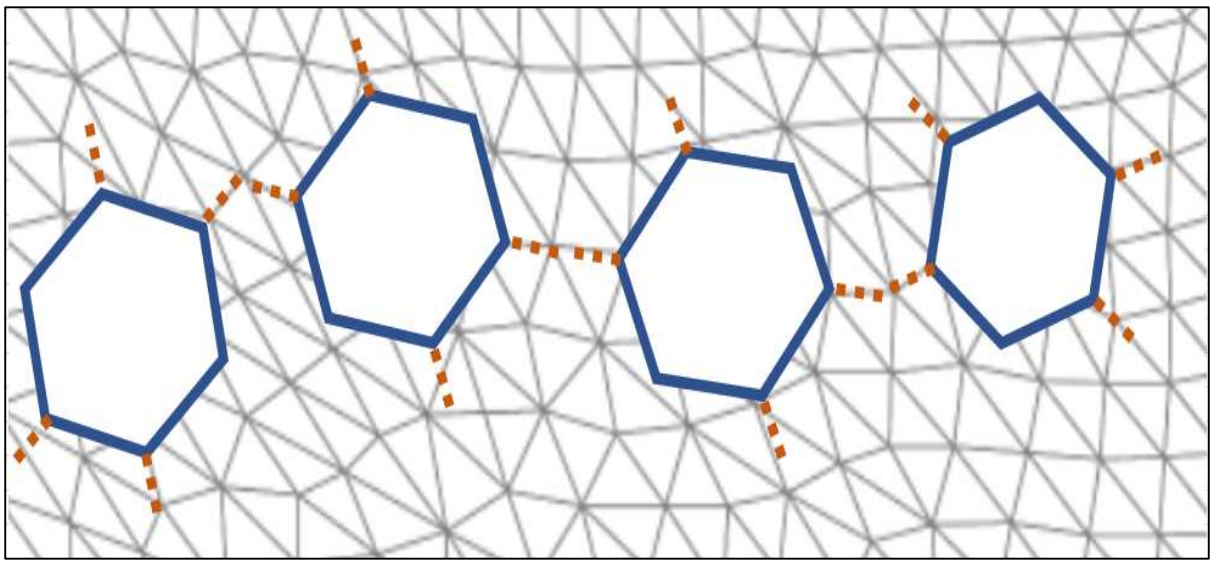

Fig. 16: Correlated clathrate hydrates distorting the tetrahedral lattice of the inhibitor solvent

The associated topology has circumambient properties such that it can envelope, isolate, and form simultaneous, non-covalent interactions with the surrounded entities [84]. Under such sterically-restricted conditions, water molecules are unable to form a full complement of closed hydrogen bonds and a cluster (either *ortho-* or *para-*) will possess multiple, dangling hydrogen bonds, thereby increasing the magnetic exchange interaction potential.

Whilst magnetic exchange interactions are conventionally established through direct and super-exchange mechanisms between metal centres or metal centres and various ligands, exchange can also occur via intermolecular hydrogen bonding interactions [10,11]. Changes in bond lengths and angles along the exchange pathway affect the hopping integrals between magnetic centres, thereby altering the magnetic exchange effect.

A clathrate hydrate devoid of guest molecules would constitute the lowest density ice phase of ice water. Low-density phases are thermodynamically favoured by low or negative pressures, so empty clathrates may be a stable crystal of water when subjected to tension or stretching [27,28]. In this scenario, the methane molecules would remain trapped within the inhibitor solvent such that any $1/\sqrt{x}$, long-range vdW interaction with their former hosts is diminished [9].

Appendix E

Lorentz rotations correspond to changes in hyperbolic surface area. The concept of rapidity [85] is commonly used as a measure of relativistic velocity and simplifies the Lorentz rotational transformation formulas. Rapidity φ is defined as the hyperbolic angle that differentiates two frames of reference in relativistic motion with each frame having time and distance coordinates. Since the distance coordinate on a hyperbolic manifold is determined by the critical exponent for correlation length $v$, then $v$ can be taken as the dimensionless group parameter of rapidity, ie. d$v$ ≡ φ.

Since tanh(φ) = v/c, the 3-dimensional result is:

$$v/c = \tanh(3 \times 0.593) = 0.945 \tag{28}$$

The Lorentz transformation for length contraction/ expansion x' then gives:

$$x' = \frac{x}{\sqrt{1 - \left(\frac{v^2}{c^2}\right)}} \tag{29}$$

such that x' = correlation length ξ = 3.05 and d$V$ = 28.4.

Since

$$\xi \sim |T - T_c|^{-v} \tag{30}$$

$|T - T_c|$ ≈ 0.15K which, in terms of self-organized criticality, represents approximately 3.5 equal steps with an average critical approach temperature of 0.15K.

The calculated overall volume correlation is (3.5 x 28.4 = 99.4) as compared to the experimentally derived value of 100 recorded in **Table A1**.